\documentclass[lettersize,journal]{IEEEtran}
\usepackage{amsmath,amsfonts}
\usepackage{algorithmic}
\usepackage{array}
\usepackage{cleveref}
\usepackage[caption=false,font=normalsize,labelfont=sf,textfont=sf]{subfig}
\usepackage{makecell}
\usepackage{diagbox}
\usepackage{textcomp}
\usepackage{stfloats}
\usepackage{url}
\usepackage{verbatim}
\usepackage{graphicx}
\allowdisplaybreaks
\usepackage{cite}
\usepackage[square, comma, sort&compress, numbers]{natbib}
\newtheorem{theorem}{Theorem}
\newtheorem{remark}{Remark}
\usepackage[ruled]{algorithm2e}
\begin{document}

\title{Knowledge-driven Meta-learning for CSI Feedback}

\author{Han Xiao, Wenqiang Tian, Wendong Liu, Jiajia Guo, Zhi Zhang, Shi Jin, Zhihua Shi, Li Guo, and Jia Shen

\thanks{H. Xiao, W. Tian, W. Liu, Z. Zhang, Z. Shi, L. Guo, and J. Shen are with the Dept. of Standardization, OPPO Research Institute, Beijing, China (e-mail: xiaohan1@oppo.com; tianwenqiang@oppo.com; liuwendong1@oppo.com; zhangzhi@oppo.com; szh@oppo.com; v-guoli@oppo.com; sj@oppo.com). 
}
\thanks{
J. Guo and S. Jin are with the National Mobile Communications
Research Laboratory, Southeast University, Nanjing 210096, China
(e-mail: jiajiaguo@seu.edu.cn, jinshi@seu.edu.cn).
}

}



\maketitle

\begin{abstract}
Accurate and effective channel state information (CSI) feedback is a key technology for massive multiple-input and multiple-output systems. Recently, deep learning (DL) has been introduced for CSI feedback enhancement through massive collected training data and lengthy training time, which is quite costly and impractical for realistic deployment. In this article, a knowledge-driven meta-learning approach is proposed, where the DL model initialized by the meta model obtained from meta training phase is able to achieve rapid convergence when facing a new scenario during target retraining phase. Specifically, instead of training with massive data collected from various scenarios, the meta task environment is constructed based on the intrinsic knowledge of spatial-frequency characteristics of CSI for meta training. Moreover, the target task dataset is also augmented by exploiting the knowledge of statistical characteristics of wireless channel, so that the DL model can achieve higher performance with small actually collected dataset and short training time. In addition, we provide analyses of rationale for the improvement yielded by the knowledge in both phases. Simulation results demonstrate the superiority of the proposed approach from the perspective of feedback performance and convergence speed.
\end{abstract}


\section{Introduction}
\IEEEPARstart{M}{assive} multiple-input multiple-output (MIMO) technology is one of the key physical layer technologies in the fifth generation (5G) system \cite{araujo2016massive,5Ghai,series2015imt,TS38913} and also part of pre-research of the sixth generation (6G) system \cite{jiang2021road,yang20226g}. To fully exploit the potential of massive MIMO system, accurate channel state information (CSI) feedback has been intensively studied for decades. Along with the standardization work in the 3rd Generation Partnership Project (3GPP), various solutions based on the Type I and enhanced Type II (eType II) codebook have been proposed to improve the CSI feedback performance \cite{TS0000,TS0001,TS0002,TS0003}. However, to resolve the issues of larger CSI feedback overhead and insufficient recovery accuracy, methods for further enhancing the CSI feedback are still being actively studied.  

Due to the successful deep learning (DL) in the field of computer vision (CV) and natural language processing (NLP), the combination of wireless communication and DL has attracted great attention in recent years. A series of creative topics have been proposed one after another such as DL-based enhancement of air interface \cite{hoydis2021toward} especially in physical layer \cite{huang2019deep,qin2019deep}, intelligent cognitive radio \cite{qin2020pathway} and semantic communication \cite{qin2023generalized}. One of the implementations of DL-based physical layer is DL-based CSI feedback \cite{9931713}
which achieves higher CSI recovery accuracy with reduced feedback overhead. A framework of autoencoder for CSI feedback, namely CsiNet \cite{wen2018deep} is first proposed, where an encoder at the user equipment (UE) compresses the channel matrix and a decoder at the base station (BS) recovers the corresponding channel matrix. Subsequently, as shown in Table \ref{citepapers}, a series of follow-up works are conducted under various conditions, which can be divided into different strategies. First, a series of works focus on novel DL neural network (NN) design to improve feedback performance including modified Transformer \cite{vaswani2017attention} backbone \cite{xiao2022ai,bi2022novel}, introducing attention mechanism \cite{cai2019attention,song2021saldr,ji2021clnet} and non-local module \cite{yu2020ds}. Second, many relevant works exploit side information to improve feedback accuracy. In \cite{li2020spatio}, decoupled spatio-temporal feature is extracted by deep recurrent NN (RNN) \cite{medsker2001recurrent}. By exploiting the partial reciprocity of the downlink and uplink channels, \cite{liu2021hyperrnn} introduces the uplink CSI into the downlink channel reconstruction to further improve the feedback performance. In \cite{liu2021evcsinet,shen2022mixernet}, principal component of CSI, i.e., eigenvectors is considered to further improve feedback efficiency. Third, some works consider the joint design of DL-based CSI feedback with other modules in wireless communication systems, including channel estimation \cite{zhao2022joint}, precoding \cite{9347820} and source-channel coding \cite{xu2022deep}. Furthermore, there are works foucus on solving the problem of practical implementation. \cite{liu2021evcsinet,shen2022mixernet} achieve direct fair performance comparison with the existing codebook based
solutions and can be well adapted to existing wireless communication systems. Scheme for  multi-user MIMO scenario is also proposed in \cite{mashhadi2020distributed}.
 Additionally, 3GPP release 18 work includes a new study item (SI), the ``Artificial Intelligence / Machine Learning for NR Interface'', and the DL-based CSI feedback is considered as one of the most promising use cases \cite{213599}, where the studies include the evaluation methodology, potential specification impacts and other possible aspects.
 
Although DL-based CSI feedback has great potential, there are still some challenges in standardization and practical deployment. For instance, the generalization issue should be considered since DL method tends to express the scenario-specific property. Furthermore, plenty of data of target scenario are always necessary for training a well-performed model, which is quite impractical for deployment due to extremely high cost. Therefore, an attractive methodology referred to as meta-learning \cite{vilalta2002perspective,vanschoren2019meta} is proposed to resolve the above challenges, which is capable of inferring the inductive bias \cite{baxter2000model} from multiple related tasks reduces the requirement of training data and time for a new target task. Specifically, model-agnostic meta-learning (MAML) \cite{finn2017model} and Reptile \cite{nichol2018first} are representative schemes, wherein the model can learn a good initialization during meta training phase and then rapidly achieve the convergence with a few of data for a new target task scenario in target retraining phase. To exploit the advantage of meta-learning, a series of meta-learning-based wireless communication schemes are proposed \cite{zeng2021downlink,tolba2021meta, park2020learning,9257198,yang2020deep}. Conventional meta-learning method \cite{zeng2021downlink,tolba2021meta} utilize meta-learning for CSI feedback, where the model is initialized in meta training phase with massive CSI samples of multiple various scenarios and then achieves a quick convergence with small dataset for a new target scenario. 

\begin{table}[t]
\centering
\caption{Related Works of DL-Based CSI Feedback}
\label{citepapers}
\setlength{\tabcolsep}{1.5mm}{
\begin{tabular}{|c|c|}
\hline
Strategies          & Related Works   \\ \hline
\makecell{Novel NN \\ design}             & \makecell{Modified Transformer
\cite{xiao2022ai,bi2022novel}\\ Attention mechanism \cite{cai2019attention,song2021saldr,ji2021clnet}  \\ Non-local module \cite{yu2020ds}} \\ \hline
\makecell{Side\\ information\\exploitation} & \makecell{Spatio-temporal feature \cite{li2020spatio} \\ Reciprocity of downlink and uplink channels \cite{liu2021hyperrnn} \\ Principal component information \cite{liu2021evcsinet,shen2022mixernet}} \\ \hline
\makecell{Joint design\\ of multiple\\ modules}        & \makecell{Joint feedback and estimation \cite{zhao2022joint,9347820} \\ Joint feedback and precoding \cite{9347820} \\ Joint feedback and source-channel coding \cite{xu2022deep}} \\ \hline
\makecell{Practical\\ application}    & \makecell{Adapt to existing communication system \cite{liu2021evcsinet,shen2022mixernet} \\ Multi-user scenario \cite{mashhadi2020distributed} \\ Generalization, data cost and training complexity \cite{zeng2021downlink,tolba2021meta}} \\ \hline
\end{tabular}}
\end{table}

However, in the above-mentioned conventional meta-learning-based CSI feedback solutions, massive collected data are still necessary for meta training. Meanwhile, the meta-trained model is retrained on the original small amount of data of the target scenario, thus it might suffer from performance loss in comparison with models training on sufficient data. Moreover, the intrinsic knowledge in the filed of wireless communication has not been fully explored in the above-mentioned works, which is quite laggard in comparison with the work of utilizing the local relation between lines and textures in image data with convolutional neural network (CNN) \cite{albawi2017understanding} in CV, and the work of evacuating the serialization characteristics of language data with RNN \cite{medsker2001recurrent} in NLP. Obviously, for meta-learning-based CSI feedback, no knowledge of intrinsic characteristics of the wireless channel are utilized. Therefore, how to explore the characteristics of wireless channel to reduce the cost of collecting amounts of data in meta training phase and then enhance the performance of model in target retraining phase remains a very interesting and challenging issue.

In this article, a novel knowledge-driven meta-learning scheme for CSI feedback is proposed. Specifically, instead of training with the large amounts of CSI data collected from different wireless scenarios in meta training phase, one constructs the dataset for meta training phase, namely \textit{meta task environment} by exploring the intrinsic knowledge of spatial-frequency characteristic of CSI eigenvector. After the DL model obtains the initialization during meta training phase, it is capable of achieving rapid convergence and high performance by retraining on the \textit{target task dataset}, which is augmented from only a small seeded dataset of the target scenario with the assistance of the knowledge of channel statistical feature. Analyzes are also provided to illustrate the rationale of improvement yielded from knowledge. Finally, the simulation results illustrate the superiority of the proposed approach from the perspective of feedback performance and convergence speed. Note that part of the work was presented at the IEEE International Conference on Communications 2023 \cite{xiao2023knowledge}. More novel content beyound the original work including intuitive motivation, algorithm improvement, detailed analyses, abundant experiments and potential standardization impact are added in this artical. The main contributions of this artical are summarized as follows.

\begin{itemize}
\item 
A knowledge-driven meta training approach for CSI feedback is proposed. In meta training phase, the model is learnt from a meta task environment which is creatively constructed based on the intrinsic knowledge of spatial-frequency feature without any actually collected data from various scenarios. 
\item 
In target retraining phase, a novel knowledge-driven data augmentation method is proposed with the assistance of statistical feature of wireless channels on a limited amount of seeded data of the target scenario.
\item 
Corresponding analyses are also provided to illustrate the rationale for the improvement yielded by the knowledge in both phases, including i) reasonability of constructing a CSI sample using spatial-frequency structure, ii) sufficiency of information in meta task environment and iii) consistency between seeded data and augmented data features.
\item 
Various kinds of simulation results are provided to demonstrate the superiority of proposed knowledge-driven meta-learning approach over the conventional counterparts from the perspective of CSI feedback performance and convergence speed of the target task retraining. These abundant simulations are performed with 3GPP link level channels, which may hopefully provide some referable insights for 3GPP discussions in the future.
\end{itemize}

The remainder of this article is organized as follows. In section \ref{System Description}, we introduce the system model and formulate the problem of meta learning to be solved for CSI feedback. In section \ref{Motivation and Knowledge-driven Meta-Learning Framework}, the motivation of this artical is first provided, then the proposed knowledge-driven meta-learning approach is introduced. In section \ref{Analysis}, we provided the analyses to demonstrate the rationale for the improvement yielded by knowledge. Numerical experiment results are provided in section \ref{Simulation Results}. Potential standardization work and prospects of the proposed approach are discussed in sections \ref{Standardization Potential and Prospects}. Final conclusions are given in  \ref{Conclusion}.

Throughout this artical, upper-case and lower-case letters denote scalars. Boldface upper-case and boldface lower-case letters denote matrices and vectors, respectively. Calligraphic upper-case letters denote sets. $\mathbf{A}(:,\mathcal{B})$ and $\mathbf{A}(\mathcal{B},:)$ denote the sub-matrices of $\mathbf{A}$ that consist of the columns and rows corresponding to vector-indices in set $\mathcal{B}$, respectively. $\mathbb{E}\{\cdot\}$ denotes the expectation and $\mathrm{Tr}\{\cdot\}$ denotes the trace of the input matrix. Hermitian matrix of $\mathbf{A}$ is denoted by $\mathbf{A}^{\rm H}$. $\mathrm{rand}(\mathcal{A}, a)$ denotes the random sampling of $a$ samples from the input set $\mathcal{A}$ without replacement. The set of real and complex numbers are denoted by $\mathbb{R}$ and $\mathbb{C}$, respectively. $|\cdot|$ denotes the cardinality of the input set or the absolute value of the input scalar.

\section{System Description}\label{System Description}
\subsection{System Model}
A typical MIMO system with $N_{\rm{t}} = N_{\rm h}N_{\rm v}$ transmitting antennas at BS and $N_{\rm{r}}$ receiving antennas at UE is considered, where $N_{\rm h}$ and $N_{\rm v}$ are the numbers of horizontal and vertical antenna ports, respectively.
Note that proposed methods are suitable for antennas with either dual or single polarization. The single polarization is considered to illustrate the basic principle in algorithm introduction for simplicity and dual polarization is applied in the simulation for practical. The downlink channel in time domain can be denoted as $\widehat{\mathbf{H}} \in \mathbb{C}^{N_{\rm r}\times N_{\rm t}\times N_{\rm d}}$, where $N_{\rm d}$ is the number of paths with various delays. By conducting Discrete Fourier transform (DFT) over the delay-dimension of $\widehat{\mathbf{H}}$, the downlink channel in frequency domain $\widetilde{\mathbf{H}}\in\mathbb{C}^{N_{\rm r}\times N_{\rm t}\times N_{\rm sc}}$ can be written as
\begin{equation}\label{Hf_1}
\widetilde{\mathbf{H}}= \big[\widetilde{\mathbf{H}}_1,\widetilde{\mathbf{H}}_2,\cdots,\widetilde{\mathbf{H}}_{N_{\rm{sc}}}\big],
\end{equation}
where $N_{\rm{sc}}$ is the number of subcarriers, and $\widetilde{\mathbf{H}}_k\in\mathbb{C}^{N_{\rm r}\times N_{\rm t}}, 1\leq k\leq N_{\rm sc}$ denotes downlink channel on the $k$th subcarrier. Furthermore, considering the limitation on the CSI feedback overhead, the whole band of $N_{\rm sc}$ subcarriers are uniformly divided into $N_{\rm sb}$ subbands, wherein the CSI eigenvector feedback is performed on each subband which consists of $N_{\rm gran}$ subcarriers with $N_{\rm sc} = N_{\rm gran}N_{\rm sb}$. For brevity, single-layer configuration for downlink transmission is assumed and the basic principle of this artical can be generalized to the case of multi-layer. The corresponding eigenvector for the $l$th subband $\mathbf{w}_l\in\mathbb{C}^{N_{\rm t}\times 1}$ with normalization $||\mathbf{w}_l||_2=1$, can be calculated by the eigenvector decomposition on the subband as
\begin{equation}
\label{eigdecomposition}
\left(\frac{1}{N_{\rm{gran}}}\sum_{k=(l-1)N_{\rm{gran}}+1}^{lN_{\rm{gran}}}\widetilde{\mathbf{H}}_k^{\rm H}\widetilde{\mathbf{H}}_k\right) \mathbf{w}_l =\lambda_l \mathbf{w}_l,
\end{equation}
where $1\leq l \leq N_{\rm sb}$ and $\lambda_l$ represents the corresponding maximum eigenvalue for the $l$-th subband. Therefore, the CSI matrix can be written as
\begin{equation}\label{Wi}
\mathbf{W}= \big[\mathbf{w}_{1},\mathbf{w}_{2},\cdots,\mathbf{w}_{N_{\rm{sb}}}\big] \in \mathbb{C}^{N_{\rm t}\times N_{\rm{sb}}},
\end{equation}
wherein total $N_{\rm{sb}}N_{\rm t}$ complex coefficients need to be compressed at the UE and then recovered at the BS side.

Generally, the optimization objective for CSI feedback can be given as
\begin{equation}\label{score_function_1}
\begin{split}
\min_{\mathfrak{F}} -\rho(\mathbf{W}, \mathbf{W}')
 = \min_{\mathfrak{F}} -\frac{1}{N_\textrm{sb}}\sum_{l=1}^{N_\textrm{sb}}\left(\frac{\|\mathbf{w}^{\rm H}\mathbf{w}'\|_2}{\|\mathbf{w}\|_2\|\mathbf{w}'\|_2}\right)^2
\end{split}
\text{,}
\end{equation}
where $\rho(\cdot,\cdot) \in [0,1]$ denotes the squared generalized cosine similarity (SGCS) which has been a widely used evaluation metric for CSI feedback. A larger $\rho$ indicating higher CSI recovery accuracy. Here, $\|\cdot\|_2$ denotes the $\ell_{2}$ norm, $\mathbf{w}_{l}$ and $\mathbf{w}'_{l}$ represent the original and recovered CSI eigenvector on the $l$-th subband, respectively. $\mathfrak{F}$ represents the alternative CSI feedback schemes such as codebook based Type I, eType II \cite{TS0001,TS0002,TS0003} and DL-based autoencoder \cite{liu2021evcsinet}.


\subsection{DL-based CSI Feedback}\label{DL-based CSI Feedback summary}
In this subsection, the DL-based CSI feedback using autoencoder is introduced, where the encoder and decoder with neural network (NN) are deployed at UE and BS, respectively. The encoder and decoder function with trainable parameters $\Theta_{\rm{E}}$ and $\Theta_{\rm{D}}$ can be denoted as $f_{\rm e}(\cdot;\Theta_{\rm{E}})$ and $f_{\rm d}(\cdot;\Theta_{\rm{D}})$, respectively, thus the DL-based autoencoder $f_{\rm a}(\cdot; \Theta)$ with trainable parameters $\Theta = \{\Theta_{\rm{E}}, \Theta_{\rm{D}}\}$ for CSI feedback can be represented as
\begin{equation}\label{autoencoder_1}
\begin{split}
\mathbf{W}' =  f_{\rm d}(f_{\rm e}(\mathbf{W};\Theta_{\rm{E}}); \Theta_{\rm{D}}) = f_{\rm a}(\mathbf{W};\Theta)
\end{split}
\text{,}
\end{equation}
where the encoder first compresses original CSI $\mathbf{W}$ and quantizes it to a bitstream $\mathbf{b}$ of length $B$. Then the decoder recover $\mathbf{W}'$ from $\mathbf{b}$. During training phase, the encoder and decoder are jointly optimized to solve (\ref{score_function_1}) with sufficient numbers of CSI eigenvector samples. In this work, we consider an offline-training-online-testing framework of CSI feedback, where the encoder and decoder are jointly trained offline and further deployed on the UE and BE, respectively.

\subsection{Meta-learning-based CSI Feedback}\label{Meta-learning based CSI Feedback}

In this subsection, the meta-learning-based CSI feedback is introduced. Generally, the goal of meta-learning based CSI feedback is to find a good initialization of $\Theta = \{\Theta_{\rm{E}}, \Theta_{\rm{D}}\}$, so that the autoencoder can quickly achieve the convergence with a small dataset and a few training steps for a new task. Specifically, the procedure of meta-learning based CSI feedback can be divided into two phases, i.e., the meta training phase and target retraining phase.

During meta training phase, the model is trained over a big dataset namely meta task environment consisting of $T$ CSI tasks of diverse scenarios, which can be defined as meta task environment $\mathcal{T}_{\rm meta} = \{\mathcal{T}_1,...,\mathcal{T}_{T}\}$, wherein each task $\mathcal{T}_j = \{\mathbf{W}^{j}_{1},...,\mathbf{W}^{j}_{|\mathcal{T}_j|}\}, 1\leq j \leq T$ consists of $|\mathcal{T}_j|$ CSI samples denoted as $\mathbf{W}^{j}_{i}, 1\leq i \leq |\mathcal{T}_j|$. Tasks can be distinguished based on different CSI  characteristics of different scenarios. For example, the CSI samples from different UEs in an urban macrocell have similar wireless characteristics, while the ones in another cell hold relatively different characteristics. Thus the datasets collected from the two cells formulate two different tasks. Based on the meta task environment $\mathcal{T}_{\rm meta}$, meta-learning algorithms such as MAML \cite{finn2017model} and Reptile \cite{nichol2018first} can be performed to learn the initial parameters $\widehat{\Theta}$ that can be described as
\begin{equation}\label{meta_training_pahse_problem}
\begin{split}
\min_{\widehat{\Theta}} \mathbb{E}_{\mathcal{T}_j \subset \mathcal{T}_{\rm meta}}\left[-\rho(\mathcal{T}_j,f_{\rm a}(\mathcal{T}_j;\mathrm{U}^g_{\mathcal{T}_j}(\widehat{\Theta})))   \right]
\end{split}
\text{,}
\end{equation}
where $\mathrm{U}^g_{\mathcal{T}_j}(\widehat{\Theta})$ is the operator that updates $\widehat{\Theta}$ for $g$ training steps using the data sampled from $\mathcal{T}_j$. The initialization $\widehat{\Theta}$ learnt in above procedure is expected to has the ability of quick adaptation with small amount of data on an unobserved target task $\mathcal{T}_{\rm target}$ from a new scenario.

Secondly, the target retraining phase can be formulated as
\begin{equation}\label{target_retraining_pahse_problem}
\begin{split}
\min_{\Phi = \mathrm{U}^g_{\mathcal{T}_{\rm target}}(\widehat{\Theta})} -\rho(\mathcal{T}_{\rm target},f_{\rm a}(\mathcal{T}_{\rm target};\Phi))
\end{split}
\text{,}
\end{equation}
where $\Phi$ denotes the possible parameter sets trained on $\mathcal{T}_{\rm target}$ after $g$ retraining steps based on the initialization $\widehat{\Theta}$, which indicates that the convergence on a target task of new scenario can be rapidly reached with only a few retraining steps.

However, the existing meta-leaning based CSI feedback still has to face two major challenges.
\begin{itemize}
\item During meta training phase, it requires sufficient CSI samples to construct the meta task environment $\mathcal{T}_{\rm meta}$ to solve (\ref{meta_training_pahse_problem}), which is extremely costly since it is impractical to collect all existing types of wireless scenarios with adequate diversity.
\item During target retraining phase, despite the rapid convergence for solving (\ref{target_retraining_pahse_problem}) using small dataset $\mathcal{T}_{\rm target}$ with initialization $\widehat{\Theta}$, it is always difficult to achieve comparable performance with using big dataset in target scenario.
\end{itemize}
To solve those two challenges, the knowledge-driven meta-learning approach consisting of knowledge-driven meta training phase and knowledge-driven target retraining phase is proposed.

\section{Knowledge-driven Meta-Learning}
\label{Motivation and Knowledge-driven Meta-Learning Framework}
\begin{figure*}[t]
\centering
\includegraphics[scale=0.68]{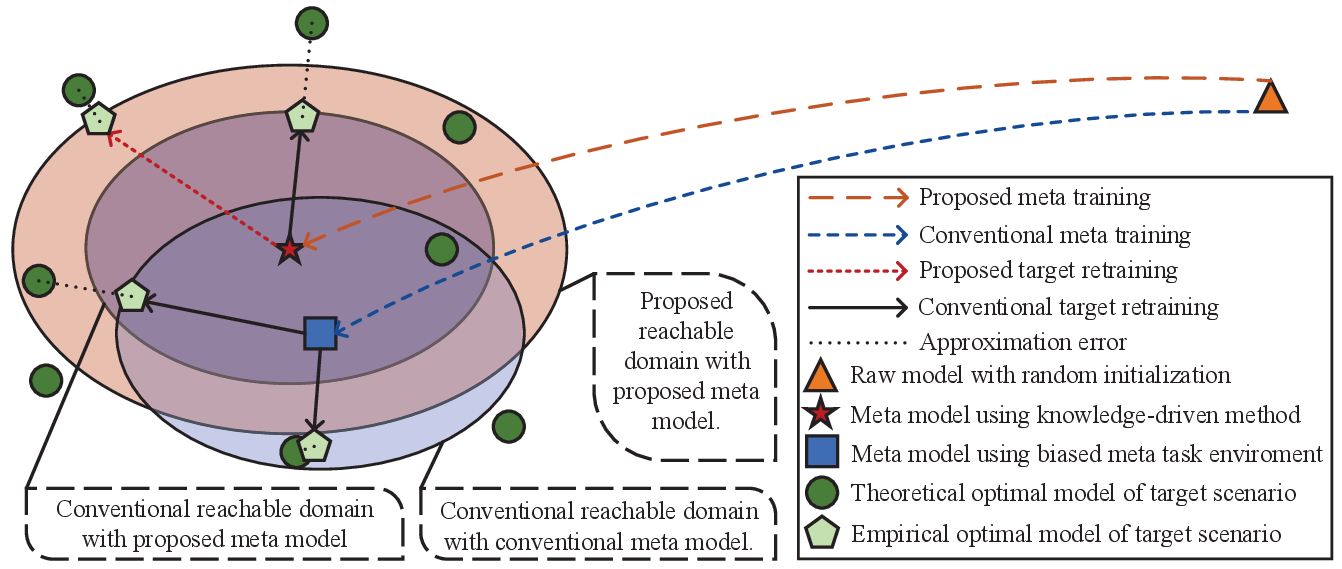}
\caption{Illustrate of motivation by intuitively comparing the different meta training and target retraining methods. For simplicity, models are assumed with two parameters thus can be depicted in two-dimensional plane. Same principle can be generalized to models with multiple parameters in hyperplane. Reachable domain is depicted as ellipse and may actually be other irregular shapes. Conventional and proposed meta model denote the meta model obtained from conventional and proposed meta training, respectively. Conventional and proposed reachable domain denote the reachable domain using conventional and proposed target retraining, respectively.}
\label{motivation}
\end{figure*}

In this section, motivation of proposed approach is first presented. Then the detailed design of knowledge-driven meta training and target retraining phases are introduced with heuristic knowledge and corresponding algorithms.

\subsection{Motivation}

In meta-learning-based CSI feedback mentioned in subsection \ref{Meta-learning based CSI Feedback}, the meta training phase is generally to train a raw randomly-initialized model to a meta model, initialized by which the one is expected to have the ability to rapidly fit different new target scenarios. This procedure can be depicted by the dashed line from the triangle to the pentagram or square in Fig. \ref{motivation}, where the pentagram and square denote the meta models using proposed and conventional methods in meta training phase, respectively, and are expected to be averagely close to the theoretical optimal target models of different scenarios represented by circles. In conventional meta-learning methods, a large amount of actually collected data with high diversity for constructing meta task environment is required in meta training phase, to guarantee that the meta model equips with sufficient knowledge of CSI and thus converges to a balanced position in Fig. \ref{motivation}. However, collecting all existing types of scenarios with adequate diversity is impractical, even simulation data such as using clustered delay line (CDL) channel models defined in 3GPP \cite{TR0004} cannot cover all possible wireless scenarios in reality. In this case the meta task environment in conventional method can be relatively biased, which leads to a biased meta model denoted by square in Fig. \ref{motivation}. A reachable domain centered on the meta model can be defined as the possible positions that can be converged to in the target retraining phase with a limited number of training steps and small target task dataset. The biased meta model is difficult to generlize to a distant theoretical optimal target model with small approximation error\footnote{Note that the approximation error (dotted line) between the theoretical optimal model (circle) and empirical optimal model (pentagon) is inevitable due to the imperfect empirical hyperparameter adjustment when model training and the error between empirical distribution of data and theoretical distribution.} due to a biased reachable domain. A performance loss occurs as a result. Therefore, to learn a relatively unbiased meta model during meta training phase, the knowledge-driven meta-learning is proposed. In Fig. \ref{motivation}, difference between two conventional reachable domains with proposed and conventional meta model imply the benefit of proposed knowledge-driven meta-learning.
\begin{figure}[t]
\centering
\includegraphics[scale=0.83]{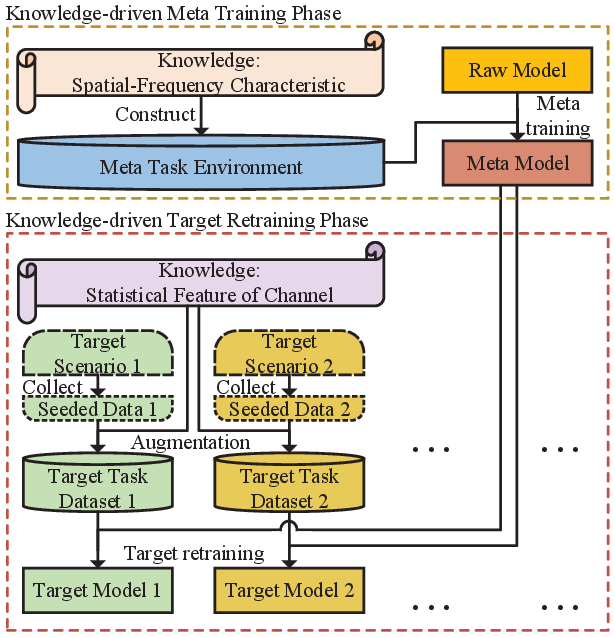}
\caption{Proposed knowledge-driven meta-learning approach for CSI feedback}
\label{pic_proposed}
\end{figure}

Since the conventional meta-learning methods utilize small target task dataset in target retraining phase, the reachable domain is limited and cannot cover the theoretical optimal solution. Therefore, the target model tends to converge to the edge of the reachable domain when theoretical optimal solution is distant and falls outside the reachable domain, leading to a large approximation error so that a performance loss. Data augmentation seeded by the small target task dataset emerge as the requirement to expand the reachable domain. However, the existing data augmentation methods such as flipping and cyclic shift \cite{xiao2022ai} borrowed from CV and NLP are not suitable for wireless channal data due to the fact that they fail to consider the communication knowledge of intrinsic feature of wireless channel. Therefore the advanced knowledge-driven target retraining phase is proposed. In Fig. \ref{motivation}, difference between conventional and proposed reachable domains with proposed  meta model imply the benefit of proposed knowledge-driven target retraining.

As shown in Fig. \ref{pic_proposed}, the proposed knowledge-driven meta-learning framework consists of i) knowledge-driven meta training phase and ii) knowledge-driven target retraining phase. Completely different from existing methods, the meta model learns from proposed relatively unbiased meta task environment which is constructed by utilizing spatial-frequency characterisic knowledge of CSI, then rapid convergence and high performance can be achieved in expanded reachable domain levering the statistical feature knowledge of channel.

\subsection{Knowledge-Driven Meta Training Phase}\label{Knowledge-dirven Meta Training Phase}
\subsubsection{Spatial-Frequency Characteristic}\label{knowledge_A}
Generally, considering the intrinsic structure of the CSI matrix, $\mathbf{W} \in \mathbb{C}^{N_{\rm t}\times N_{\rm{sb}}}$ can be decomposed as
\begin{equation}\label{decompose}
\mathbf{W} = \mathbf{S}\mathbf{E}\mathbf{F}^{\rm H}
\end{equation}
where $\mathbf{S}\in \mathbb{C}^{N_{\rm t} \times N_{\rm t}}$ is constructed with $N_{\rm t}$ orthogonal basis vectors in spatial domain and $\mathbf{F} \in \mathbb{C}^{N_{\rm sb} \times N_{\rm sb}}$ is constructed with $N_{\rm sb}$ orthogonal basis vectors in frequency domain. Specifically, both $\mathbf{S}$ and $\mathbf{F}$ are unitary matrices, which indicate the full-rank spatial-frequency characteristic. The projection coefficient matrix $\mathbf{E} \in \mathbb{C}^{N_{\rm t} \times N_{\rm sb}}$ represents that each CSI eigenvector $\mathbf{W}$ can be completely expressed by the linear combination of the orthogonal basis vectors in $\mathbf{S}$ and $\mathbf{F}$. Obviously, the distribution of the elements in $\mathbf{E}$ with relatively larger amplitude determines the dominant spatial-frequency feature of $\mathbf{W}$ given the same $\mathbf{S}$ and $\mathbf{F}$, where the dominant spatial-frequency features can be considered as the intrinsic
knowledge and hence can be learnt by the DL model during the meta-training phase.

\subsubsection{Knowledge-Driven Meta Training}
Inspired by the intrinsic knowledge of spatial-frequency feature, a knowledge-driven algorithm is proposed to solve (\ref{meta_training_pahse_problem}). During the meta training phase, the meta task environment $\mathcal{T}_{\rm meta} = \{\mathcal{T}_1,...,\mathcal{T}_{T}\}$ consisting of $T$ tasks is firstly established, where the construction approach of CSI matrix in each task explores the CSI decomposition formula in section \ref{knowledge_A}. Each task usually consists of different CSI samples from specific number of UEs that can be sampled on various number of slots. Specifically, denote $N_{\rm ue\it, j}$ and $N_{\rm slot\it, j}$ as the number of UEs and slots for the $j$-th task $\mathcal{T}_j$, $1\leq j \leq T$, respectively, which can be set as
\begin{equation}\label{random_ue}
\begin{split}
N_{\rm ue,\it j} = \mathrm{rand}(\{1,...,\widehat{N}_{\rm ue}\}, 1)
\end{split}
\text{,}
\end{equation}
\begin{equation}\label{random_slot}
\begin{split}
N_{\rm slot,\it j} = \mathrm{rand}(\{1,...,\widehat{N}_{\rm slot}\}, 1)
\end{split}
\text{,}
\end{equation}
where $\mathrm{rand}(\mathcal{A}, a)$ denotes the random sampling of $a$ samples from the input set $\mathcal{A}$ without replacement, $N_{\rm slot,\it j}N_{\rm ue,\it j} = |\mathcal{T}_j|$, $\widehat{N}_{\rm ue}$ and $\widehat{N}_{\rm slot}$ denote the maximum number of UEs and maximum number of slots of CSI that can be generated in one task, respectively.

Moreover, according to the intrinsic knowledge of spatial-frequency feature, to generate the CSI samples in the $j$-th task $\mathcal{T}_j$, $P$ groups of spatial orthogonal basis vector and one group of frequency orthogonal basis vector can be firstly provided as
\begin{equation}\label{special_basis}
\begin{split}
\mathbf{S}_{p} = [\mathbf{s}_{p,1},...,\mathbf{s}_{p,N_{\rm t}}] \in \mathbb{C}^{N_{\rm t} \times N_{\rm t}}, 1\leq p \leq P
\end{split}
\text{,}
\end{equation}
\begin{equation}\label{frequency_basis}
\begin{split}
\mathbf{F} = [\mathbf{f}_{1},...,\mathbf{f}_{N_{\rm sb}}] \in \mathbb{C}^{N_{\rm sb} \times N_{\rm sb}}
\end{split}
\text{,}
\end{equation}
respectively, where each column of $\mathbf{S}_{p}$ and $\mathbf{F}$ is an orthogonal basis vector. Specifically, each basis in $\mathbf{S}_p$ indicates a beam direction in spatial domain, and multiple groups of orthogonal basis vectors are designed in order to improve the diversity of spatial features. Moreover, we propose a class of feasible orthogonal basis construction methods with same effectiveness, including singular value decomposition (SVD) based, Schmidt orthogonalization (SMT) based and discrete fourier transform (DFT) based methods, where the DFT based method is introduced in subsection \ref{methods_formulating_basis}, and SVD and SMT based ones are depicted in Appendix \ref{SVD_method} and \ref{SMT based basis formulation}, respectively.

\begin{algorithm}[t]
\caption{Knowledge-driven Meta Training Phase}
\label{alg_1}
\textbf{Initialization}:$\widehat{N}_{\rm ue}$, $\widehat{N}_{\rm slot}$, $T$, $\alpha$, $\beta$, $g$, $\epsilon$, $\widehat{\Theta}$\;
Formulate the feature basis using (\ref{special_basis}) to (\ref{frequency_basis})\;
\For{$j = 1,\ldots,T$}{
    Construct structure of task $\mathcal{T}_{j}$ using (\ref{random_ue}), (\ref{random_slot}) and (\ref{random_group}-\ref{random_f_0})\;
    \For{$m = 1,\ldots,N_{\rm ue, \it j}$}{
    Construct structure of UE $m$ using (\ref{random_s_1}-\ref{random_M_m})\;
    \For{$n = 1,\ldots,N_{\rm slot, \it j}$}{
    Construct CSI of slot $n$ using (\ref{random_s_2}-\ref{csi_norm_1})\;
	}
	}
Update model parameters using (\ref{meta_training_1}).
}
\end{algorithm}

Next, the method of generating CSI samples for the $j$-th task $\mathcal{T}_j$ is introduced. The group index $p_j$ for task $\mathcal{T}_j$ are first randomized by
\begin{equation}\label{random_group}
\begin{split}
\{p_j\} = \mathrm{rand}(\{1,...,P\}, 1)
\end{split}
\text{,}
\end{equation}
and the indices of dominant spatial and frequency feature vectors are also randomized by
\begin{equation}\label{random_s_0}
\begin{split}
\widehat{\mathcal{S}}_{j} = \mathrm{rand}(\{1,...,N_{\rm t}\}, L_{\rm task})
\end{split}
\text{,}
\end{equation}
\begin{equation}\label{random_f_0}
\begin{split}
\widehat{\mathcal{F}}_{j} = \mathrm{rand}(\{1,...,N_{\rm sb}\}, M_{\rm task})
\end{split}
\text{,}
\end{equation}
respectively, where the parameters $L_{\rm task} \leq N_{\rm t}$ and $M_{\rm task} \leq N_{\rm sb}$ are defined to constrain the degree of feature diversity of the task in spatial and frequency domain, respectively.

For the $m$-th UE $1\leq m \leq N_{\rm ue}$ in task $\mathcal{T}_j$, the indices of the dominant spatial and frequency feature vectors are also randomized by
\begin{equation}\label{random_s_1}
\begin{split}
\widetilde{\mathcal{S}}_{m} = \mathrm{rand}(\widehat{\mathcal{S}}_{j}, L_m)
\end{split}
\text{,}
\end{equation}
\begin{equation}\label{random_f_1}
\begin{split}
\widetilde{\mathcal{F}}_{m} = \mathrm{rand}(\widehat{\mathcal{F}}_{j}, M_m)
\end{split}
\text{,}
\end{equation}
respectively, where the degree of feature diversity in spatial and frequency domain $L_{\rm m}$ and $M_{\rm m}$ are both UE-specific and defined as
\begin{equation}\label{random_L_m}
\begin{split}
\{L_{m}\} = \mathrm{rand}(\{1,...,L_{\rm task}\}, 1)
\end{split}
\text{,}
\end{equation}
\begin{equation}\label{random_M_m}
\begin{split}
\{M_{m}\} = \mathrm{rand}(\{1,...,M_{\rm task}\}, 1)
\end{split}
\text{,}
\end{equation}
respectively.

Similarly for the $n$-th slot $1\leq n \leq N_{\rm slot}$ of the $m$-th UE in task $\mathcal{T}_j$, the spatial and frequency dominant vectors are respectively selected from corresponding dominant vectors of the UE, thus the feature is maintained for the $m$-th UE but distinguished between different slots, i.e.,
\begin{equation}\label{random_s_2}
\begin{split}
\mathcal{S}_{m,n} = \mathrm{rand}(\widetilde{\mathcal{S}}_{m}, \lceil\alpha L_m\rceil)
\end{split}
\text{,}
\end{equation}
\begin{equation}\label{random_f_2}
\begin{split}
\mathcal{F}_{m,n} = \mathrm{rand}(\widetilde{\mathcal{F}}_{m},  \lceil \beta M_m \rceil)
\end{split}
\text{,}
\end{equation}
where parameters $\alpha \in (0,1]$ and $\beta \in (0,1]$ are set to scale the diversity of feature of each slot. Consequently, a CSI sample for the $n$-th slot of the $m$-th UE in task $\mathcal{T}_j$ can be generated as
\begin{equation}\label{csi_gen_1}
\begin{split}
\mathbf{W}^{j}_{m,n} = \mathbf{S}_{p_j}(:,\mathcal{S}_{m,n}) \widehat{\mathbf{E}}\mathbf{F}^{\rm H}(:,\mathcal{F}_{m,n})
\end{split}
\text{,}
\end{equation}
where the elements in $\widehat{\mathbf{E}} \in \mathbb{C}^{|\mathcal{S}_{m,n}| \times |\mathcal{F}_{m,n}|}$ are independently sampled from complex normal distribution $\mathcal{CN}(0,1)$. Subband-level normalization should be performed for $1\leq l\leq N_{\rm sb}$ using
\begin{equation}\label{csi_norm_1}
\begin{split}
\mathbf{W}^{j}_{m,n}(:,l) = \frac{\mathbf{W}^{j}_{m,n}(:,l)}{||\mathbf{W}^{j}_{m,n}(:,l)||_2}
\end{split}
\text{.}
\end{equation}
Through the procedure of (\ref{random_ue}) to (\ref{csi_norm_1}) for generating each CSI sample of each UE, the meta task environment $\mathcal{T}_{\rm meta}$ can be finally constructed.

Utilizing the meta task environment $\mathcal{T}_{\rm meta}$, the meta training procedure can be conducted to solve (\ref{meta_training_pahse_problem}). The parameters of the DL model of CSI feedback is randomly initialized by $\widehat{\Theta}$. For the $j$-th task $\mathcal{T}_j$ in the meta task environment $\mathcal{T}_{\rm meta}$, $\widehat{\Theta}$ can be updated with
\begin{equation}\label{meta_training_1}
\begin{split}
\widehat{\Theta} = \widehat{\Theta} + \epsilon (\mathrm{U}^g_{\mathcal{T}_j}(\widehat{\Theta}) - \widehat{\Theta})
\end{split}
\text{,}
\end{equation}
where $\mathrm{U}^g_{\mathcal{T}_j}(\widehat{\Theta})$ is the operator that updates $\widehat{\Theta}$ for $g$ training steps on task $\mathcal{T}_j$, and $\epsilon$ denotes the step size of meta training. After that, the obtained $\widehat{\Theta}$ can be utilized as initialization for further fast retraining on a new target task of scenario. The proposed algorithm for knowledge-driven meta training phase is summarized in Algorithm \ref{alg_1}.

\subsubsection{Method of Formulating Spatial and Frequency Basis Vector Group}\label{methods_formulating_basis}

The DFT steering vectors with oversampling factors $O_{\rm h}$ and $O_{\rm v}$ corresponding to horizontal and vertical antenna ports can be expressed as.
\begin{equation}\label{app_DFT_dft_2}
\begin{split}
\mathbf{a}^{\rm h}_x = [1,...,e^{j2\pi\frac{(N_{\rm h}-1)x}{N_{\rm h}O_{\rm h}}}]^{\rm T}, 0\leq x \leq N_{\rm h}O_{\rm h}-1
\end{split}
\text{,}
\end{equation}
\begin{equation}\label{app_DFT_dft_3}
\begin{split}
\mathbf{a}^{\rm v}_y = [1,...,e^{j2\pi\frac{(N_{\rm v}-1)y}{N_{\rm v}O_{\rm v}}}]^{\rm T}, 0\leq y \leq N_{\rm v}O_{\rm v}-1
\end{split}
\text{,}
\end{equation}
Then, the spatial basis vector can be obtained using kronecker product as
\begin{equation}\label{app_DFT_dft_1}
\begin{split}
\mathbf{a}_{x,y} = \mathbf{a}^{\rm h}_x \otimes \mathbf{a}^{\rm v}_y
\end{split}
\text{,}
\end{equation}

It should be noted that the oversampling factors $O_h$ and $O_v$ are introduced to construct $P = O_{\rm h}O_{\rm v}$ groups of orthogonal basis vectors with larger diversity. Specifically, all $\mathbf{a}_{x,y}$ can be divided into $P$ groups, where each group of spatial orthogonal basis vector $\mathbf{S}_p$ consists of $N_{\rm h}N_{\rm v}$ orthogonal basis vectors with $x \in \{o_{\rm h},o_{\rm h}+O_{\rm h},\dots,o_{\rm h}+N_{\rm h}O_{\rm h} - O_{\rm h}\}$ and $y \in \{o_{\rm v},o_{\rm v}+O_{\rm v},\dots,o_{\rm v}+N_{\rm v}O_{\rm v} - O_{\rm v}\}$, where specific values $0 \leq o_h < O_h$ and $0 \leq o_v < O_v$ are fixed to define a group, respectively.

Similarly, the frequency domain basis vector group $\mathbf{F}$ can be also constructed using DFT steering vectors by
\begin{equation}\label{app_DFT_dft_4}
\begin{split}
\mathbf{F} = [\mathbf{a}_{1},...,\mathbf{a}_{N_{\rm sb}}]
\end{split}
\text{,}
\end{equation}
\begin{equation}\label{app_DFT_dft_5}
\begin{split}
\mathbf{a}_l = [1,...,e^{j2\pi\frac{(N_{\rm sb}-1)(l-1)}{N_{\rm sb}}}]^{\rm T}, 1\leq l \leq N_{\rm sb}
\end{split}
\text{.}
\end{equation}
In addition to the above DFT based method, we also propose two feasible orthogonal basis construction methods with same effectiveness, i.e., the SVD and SMT based metohds, which are depicted in Appendix \ref{SVD_method} and \ref{SMT based basis formulation}, respectively.

\subsection{Knowledge-Driven Target Retraining Phase}
\label{Knowledge-dirven Target Retraing Phase}
\begin{algorithm}[t]
\caption{Knowledge-driven Target Retraining Phase}
\label{alg_2}
\textbf{Initialization}: $\widehat{\mathcal{H}} = \{\mathcal{H}_1,\dots,\mathcal{H}_f\}$, $N_{\rm aug}$, $g'$\;

\For{$f = 1,\ldots,F$}{
\For{$q = 1,\ldots,\widetilde{N}_{\rm ue}$}{
    \For{$d = 1,\ldots,N_{\rm delay}$}{
	Calculate statistical power using (\ref{cal_DPS})\;
     Construct channel using (\ref{R_tx}-\ref{H_aug})\;
	}
Construct CSI using (\ref{Hf_1}-\ref{Wi})\;
}
Update model parameters using (\ref{target_retraining_1}).
}
\end{algorithm}

\subsubsection{Statistical Feature of Channel}\label{knowledge_B}
In this part, the statistical features of the channel in both spatial domain and time delay domain are explored. Consider there are $F$ scenarios to be deployed. Specifically, for a specific UE in one target scenario\footnote{Corner mark of indices of target scenarios is omitted for simplicity.}, denote the actually collected $\widetilde{N}_{\rm slot}$ channel samples in time domain as $\mathcal{H} = \{\widehat{\mathbf{H}}_1,...,\widehat{\mathbf{H}}_{\widetilde{N}_{\rm slot}}\}$, where each channel sample $\widehat{\mathbf{H}}_t \in \mathbb{C}^{N_{\rm r}\times N_{\rm t}\times N_{\rm d}}$, $1\leq t \leq \widetilde{N}_{\rm slot}$.

Firstly, the statistical feature in delay domain can be described by the power-delay spectrum. Denote $\widehat{\mathbf{H}}'_{t,d} \in \mathbb{C}^{N_{\rm r}\times N_{\rm t}}, 1\leq d \leq N_{\rm d}$ as the $d$-th delay of the $t$-th channel sample $\widehat{\mathbf{H}}_t$, the power of the $d$-th delay can be calculated as
\begin{equation}\label{cal_DPS}
\begin{split}
\hat{p}_d = \frac{1}{N_{\rm t}N_{\rm r}\widetilde{N}_{\rm slot}}\sum_{t=1}^{\widetilde{N}_{\rm slot}}||\widehat{\mathbf{H}}'_{t,d}||_{\rm F}^2
\end{split}
\text{,}
\end{equation}

Secondly, the statistical feature in spatial domain can be demonstrated by the normalized averaged covariance matrices of the transmitting and receiving antenna ports, which can be calculated as
\begin{equation}\label{R_tx}
\begin{split}
\mathbf{R}^{\rm tx}_{d} = \frac{N_{\rm t}\sum_{t=1}^{\widetilde{N}_{\rm slot}}\widehat{\mathbf{H}}'^{\rm H}_{t,d}\widehat{\mathbf{H}}'_{t,d}}{\mathrm{Tr}(\sum_{t=1}^{\widetilde{N}_{\rm slot}}\widehat{\mathbf{H}}'^{\rm H}_{t,d}\widehat{\mathbf{H}}'_{t,d})}
\end{split}
\text{,}
\end{equation}
\begin{equation}\label{R_rx}
\begin{split}
\mathbf{R}^{\rm rx}_{d} = \frac{N_{\rm r}\sum_{t=1}^{\widetilde{N}_{\rm slot}}\widehat{\mathbf{H}}'_{t,d}\widehat{\mathbf{H}}'^{\rm H}_{t,d}}{\mathrm{Tr}(\sum_{t=1}^{\widetilde{N}_{\rm slot}}\widehat{\mathbf{H}}'_{t,d}\widehat{\mathbf{H}}'^{\rm H}_{t,d})}
\end{split}
\text{,}
\end{equation}
respectively, where the trace operation $\mathrm{Tr}(\cdot)$ is performed for normalization. Then the kronecker product is implemented on the transmitting and receiving normalized averaged covariance matrices to obtain the joint spatial feature as 
\begin{equation}\label{R_all}
\begin{split}
\mathbf{R}_{d} = \mathbf{R}^{\rm rx}_{d}\otimes\mathbf{R}^{\rm tx}_{d} \in \mathbb{C}^{N_{\rm t}N_{\rm r}\times N_{\rm t}N_{\rm r}}
\end{split}
\text{,}
\end{equation}

In traditional communication, the covariance matrix is indispensable in the widely-used linear minimum mean square error (LMMSE) channel estimation method for well representing the characteristics of the full channel. Here, by levering both $\hat{p}_d$ and $\mathbf{R}_{d}$, intrinsic feature of a wireless channel in delay domain and spatial domain can be well explored. It should be noted that the dataset of the target scenario could be very small, and thus it is not sufficient for training autoencoder with superior CSI feedback and to ensure recovery performance, even though it is able to converge quickly based on the initialization $\widehat{\Theta}$ obtained by meta training phase. Therefore, it is necessary to consider a
data augmentation seeded by $\mathcal{H}$ exploiting the  knowledge of statistical features in spatial and delay domain.

\subsubsection{Knowledge-Driven Target Retraining}
Inspired by the knowledge of statistical features in spatial and delay domain, knowledge-driven target retraining is proposed to solve (\ref{target_retraining_pahse_problem}).
Firstly, SVD is performed on $\mathbf{R}_d$, i.e.,
\begin{equation}\label{R_svd}
\begin{split}
\mathbf{U}_{d},\mathbf{D}_{d},\mathbf{V}_{d} = \mathrm{svd}(\mathbf{R}_{d})
\end{split}
\text{,}
\end{equation}
where $\mathbf{V}_d = \mathbf{U}_d^{\rm H}$ because of $\mathbf{R}_d = \mathbf{R}_d^{\rm H}$.

Secondly, the augmented channel sample for the $d$-th delay can be generated by conducting
\begin{equation}\label{h_aug}
\begin{split}
\hat{\mathbf{h}}^{\rm aug}_{d} = \sqrt{\hat{p}_{d}}\mathbf{U}_{d}\mathbf{D}_{d}^{\frac{1}{2}}\mathbf{n}
\end{split}
\text{,}
\end{equation}
where the random vector $\mathbf{n} \in \mathbb{C}^{N_{\rm t}N_{\rm r} \times 1}\sim \mathcal{CN}(0,1)$. It is proved in Appendix \ref{app_3} that $\hat{\mathbf{h}}^{\rm aug}_{d}$ satisfy the limitation of (\ref{R_d}). 

Next, $\hat{\mathbf{h}}^{\rm aug}_{d}$ can be reshaped as the channel matrix $\widehat{\mathbf{H}}^{\rm aug}_{d} \in \mathbb{C}^{N_{\rm r} \times N_{\rm t}}$. By concatenating all $N_{\rm d}$ augmented channel matrices, the augmented channel sample can be obtained as
\begin{equation}\label{H_aug}
\begin{split}
\mathbf{H}^{\rm aug} = [\widehat{\mathbf{H}}^{\rm aug}_{1},...,\widehat{\mathbf{H}}^{\rm aug}_{N_{\rm d}}]
\end{split}
\text{,}
\end{equation}
where $\mathbf{H}^{\rm aug} \in \mathbb{C}^{N_{\rm r} \times N_{\rm t} \times N_{\rm d}}$. Then the augmented CSI eigenvector sample $\mathbf{W}^{\rm aug}$ can be finally obtained by implementing (\ref{Hf_1}) to (\ref{Wi}) on $\mathbf{H}^{\rm aug}$.

For each UE, the total $N_{\rm aug}$ channel samples can be provided with $N_{\rm aug}$ randomly generated vectors $\mathbf{n}$. Moreover, for $\widetilde{N}_{\rm ue}$ UEs, we can generate totally $\widetilde{N}_{\rm ue}N_{\rm aug}$ augmented CSI eigenvector samples that can be used to construct the target task dataset $\mathcal{T}^{\rm aug}_{\rm target}$.

Based on the target task dataset $\mathcal{T}^{\rm aug}_{\rm target}$ and the initialization $\widehat{\Theta}$ obtained in knowledge-driven meta training phase, (\ref{target_retraining_pahse_problem}) can be solved with higher SGCS using a few training steps, i.e.,
\begin{equation}\label{target_retraining_1}
\begin{split}
\Phi = \mathrm{U}^{g}_{\mathcal{T}_{\rm target}}(\widehat{\Theta})
\end{split}
\text{.}
\end{equation}
The proposed algorithm for knowledge-driven target retraining phase can be summarized in Algorithm \ref{alg_2}.

\section{Analysis}\label{Analysis}
In this section, analyses are provided to illustrate the rationale for the improvement yielded by the knowledge in both phases, including reasonability of constructing CSI sample using spatial-frequency structure, sufficiency of information in meta task environment and consistency between seeded data and augmented data features.

\subsection{Reasonability of Using Spatial-Frequency Characteristic}
The proposed knowledge-driven meta training phase utilizes parts of uniform orthogonal basis vectors in the frequency and spatial domain to construct a CSI matrix. Actually, the partial basis vectors utilized can be regarded as the spatial-frequency eigenvectors of a CSI matrix and one can construct any possible CSI matrices by combining different frequency and spatial basis, which can be explained in the following Theorem. 

\begin{theorem}\label{theorem_1}
Given the small positive threshold value $\sigma>0$, the support set in spatial domain $\mathcal{S}$ and the support set in frequency domain $\mathcal{F}$, which subjects to $|\mathbf{E}(\{v\},\{l\})| \leq \sigma$, $\forall$ $v \notin \mathcal{S}$ or $l \notin \mathcal{F}$, the dominant approximation of a possible $\mathbf{W}$ expressed by $\widehat{\mathbf{W}} = \mathbf{S}\mathbf{E}'\mathbf{F}^{\rm H}$ satisfies
\begin{equation}
\begin{split}
1 - \rho(\mathbf{W},\widehat{\mathbf{W}}) \leq \frac{2}{N_\textrm{sb}}\bar{S}\bar{F}\sigma^2
\end{split}
\text{.}
\end{equation}
where $\bar{S} = N_{\rm t}-|\mathcal{S}|$, $\bar{F} = N_{\rm sb}-|\mathcal{F}|$, $\mathbf{E}'(\mathcal{S},\mathcal{F}) = \mathbf{E}(\mathcal{S},\mathcal{F})$, and $\mathbf{E}'(\{\bar{v}\},\{\bar{l}\}) = 0$, $\forall$ $\bar{v} \notin \mathcal{S} $ or $ \bar{l} \notin \mathcal{F}$.
\end{theorem}
The detailed proof is given in Appendix \ref{app_1}. Theorem \ref{theorem_1} demonstrates the support sets in spatial and frequency domain with most of energies in $\mathbf{E}$ can represent the dominant spatial-frequency features of $\mathbf{W}$ with a very small loss on SGCS. In other words, any possible CSI matrix $\mathbf{W}$ can be approximately constructed as $\widehat{\mathbf{W}}$ with only a subset of spatial and frequency basis vectors in support set $\mathcal{S}$ and $\mathcal{F}$. Since the CSI samples of a UE or a task have similar spatial-frequency characteristics, the same dominant basis vectors can be considered as the shared knowledge for construct the CSI for one UE or one task. 

Based on the above analysis, we further discuss the influence of several related parameters. $P$ in (\ref{random_group}) denote the number of groups of spatial orthogonal basis vector. Obviously, the density of total spatial basis vectors increases as $P$ increases. Therefore $P>1$ groups is set aiming to meet the requirement of more diverse spatial features in the CSI matrix. Moreover, $L_{\rm task}$ in (\ref{random_s_0}) and $M_{\rm task}$ in (\ref{random_f_0}) affect the number of basis vectors used for each task, thus constrain the degree of feature diversity of the task in spatial and frequency domain, respectively. As $L_{\rm task}$ and $M_{\rm task}$ increase, the feature diversity of each task also increases, but the  feature variance acrossing tasks decreases due to the greater possibility of overlapping basis vectors used by different tasks. Therefore, in order to ensure sufficient intra-task feature diversity and inter-task difference, it is reasonable to assign moderate values to $L_{\rm task}$ and $M_{\rm task}$.

\subsection{Sufficiency of Information in Meta Task Environment}
In this subsection, the reasonability of the proposed construction method for entire meta task environment is further discussed by levering information theory, which can be a good tool and widely used to analyze the meta-learning procedure and meta tasks \cite{yin2019meta,yao2021meta}.
\begin{remark}\label{theorem_2}
Given a task subset $\mathcal{T_{\rm sub}} = \{\mathcal{T}_{1},...,\mathcal{T}_{S}\} \subset \mathcal{T}_{\rm meta}$, arbitrary new task $\mathcal{T}_{S+1} \in \mathcal{T}_{\rm meta}$, wherein the UEs utilize different spatial and frequency basis vectors from the UEs of the tasks in $\mathcal{T_{\rm sub}}$, is represented as $\mathcal{T}_{S+1} = g(\mathcal{T_{\rm sub}},\mathcal{S}_{S+1},\mathcal{F}_{S+1})$, where $g(\cdot)$ denotes the mapping of fixing the projection coefficient matrix of a task in $\mathcal{T}_{\rm sub}$ and replacing the basis vectors with new ones indexed by $\mathcal{S}_{S+1}$ and $\mathcal{F}_{S+1}$. The following equation holds,
\begin{equation}
\label{information entropy equation}
\begin{split}
I(\widehat{\Theta} ; \mathcal{T}_{S+1} | \mathcal{T}_{\rm sub}) = H(\widehat{\Theta}) - H(\widehat{\Theta} | \mathcal{S}_{S+1}, \mathcal{F}_{S+1})
\end{split}
\text{,}
\end{equation}
\end{remark}
where $I(\cdot)$ and $H(\cdot)$ denote the mutual information and information entropy, respectively. The detailed proof of Eq. (\ref{information entropy equation}) is given in Appendix \ref{app_2}. Remark \ref{theorem_2} demonstrates that a task using different spatial-frequency basis vectors is capable of reducing the uncertainty of the $\widehat{\Theta}$. In other words, it contributes additional knowledge of spatial-frequency features to obtaining a better initialization $\widehat{\Theta}$. By constructing $P$ groups of unitary matrices $\mathbf{S}$ and $\mathbf{F}$ with full-rank structure, the meta task environments can provide sufficient and diverse intrinsic knowledge of spatial-frequency features for better meta training.

\subsection{Consistency between Seeded Data and Augmented Data Features.}
As explained in subsection \ref{Knowledge-dirven Target Retraing Phase}, the intrinsic knowledge of statistical features of the channel for a specific UE can be completely described by $\hat{p}_d$ and $\mathbf{R}_d, 1\leq d \leq N_{\rm d}$. By using proposed method in target retraining phase, the augmented channel for the $d$-th delay $\hat{\mathbf{h}}_d^{\rm aug}$  satisfy
\begin{equation}\label{R_d}
\begin{split}
\mathbb{E} [ \hat{\mathbf{h}}^{\rm aug}_d (\hat{\mathbf{h}}^{\rm aug}_d)^{\rm H} ]  = \hat{p}_d \mathbf{R}_d
\end{split}
\text{,}
\end{equation}
where $\mathbb{E}(\cdot)$ denotes the expectation operation. Prove of Eq. (\ref{h_aug}) satisfying Eq. (\ref{R_d}) is given in Appendix \ref{app_3}. Obviously, the Eq. (\ref{R_d}) holds so that the augmented and seeded data have the same statistical covariance matrix, thus statistical features of the both are aligned.

\section{Simulation Results}\label{Simulation Results}
In this section, we provide the simulation results to illustrate the superiority of the proposed knowledge-driven meta-learning scheme. Different knowledge-driven schemes in meta training phase (DFT-*, SVD-* and SMT-*) and target retraing phase (*-Aug) are evaluated, where `None' denotes no meta training phase or no knowledge-driven schemes in target retraining phase, respectively. It should be noted that the meta training phase for each basis vector scheme (DFT-*, SVD-* and SMT-*) is performed only once, where the meta models can be reused for different target retraining phase. The basic simulation parameters are listed in Table \ref{tabSystemParameters}.

In order to simulate the propagation environment of the real channel, CDL model designed for 3D channel is considered in the simulation. In addition to the two parameters of delay and power, the CDL model adds the azimuth angle of departure, azimuth angle of arrival, zenith angle of departure, and zenith angle of arrival to better characterize the spatial characteristics of the channel model. Moreover, CDL-A/C are designed for NLOS scenarios, and delay spread of 30 and 300ns corresponds to the defination of short and long delay spread \cite{TR0004}. These are in line with real complex propagation scenarios. Therefore, CDL-A with delay spread 300 ns (CDL-A300), and CDL-C with delay spread 30 and 300 ns (CDL-C30 and CDL-C300)  channel models are considered in the simulation. Here we consider three baselines, i.e., 
the i) traditional eType II codebook (eTypeII), ii) DL-based CSI feedback in subsection \ref{DL-based CSI Feedback summary} without meta training phase and knowledge enhancement in target retraing phase (None-None), and iii) state-of-art meta-learning method for CSI feedback \cite{zeng2021downlink} (MixI-None and MixII-None), where MixI-* and MixII-* denote the MAML algorithm on meta task enviroment wherein 8000 tasks are constructed by randomly selecting 16 UEs with 16 slots per task from average mixed dataset of \{CDL-A300, CDL-C30\} and \{CDL-C300 , CDL-C30\}, respectively, which brings high cost of data collection but is a fair comparison with proposed method. Moreover, the Transformer backbone for CSI feedback namely EVCsiNet-T \cite{xiao2022ai} with embedding dimension of 384, 8 heads and 6 basic blocks is implemented in evaluation. The number of feedback bits $B=64$ and the number of quantization bits $B_{\rm q}=2$ unless otherwise specified.

\begin{table}[t]
\centering
\caption{Basic simulation parameters}
\label{tabSystemParameters}
\setlength{\tabcolsep}{2mm}{
\begin{tabular}{|c|c|}
\hline
 Parameter  & Value \\
 \hline
System bandwidth &  10MHz\\
 \hline
Carrier frequency &  3.5GHz\\
 \hline
Subcarrier spacing  &  15KHz\\
 \hline
Subcarriers number $N_{\rm sc}$  &  624\\
 \hline
Subband granularity $N_{\rm gran}$  &  48\\
 \hline
Subband number $N_\textrm{\rm sb}$  &  13\\
 \hline
Horizontal Tx antenna ports per polarization $N_{\rm h}$ & 8\\
 \hline
Vertical Tx antenna ports per polarization $N_{\rm v}$ & 2\\
 \hline
Polarization & Dual\\
\hline
Tx antenna ports $N_{\rm t}$ & 32 \\
 \hline
Rx antennas $N_{\rm r}$ & 4\\
 \hline
 Channel model& CDL-A \& CDL-C\\
 \hline
 UE speed & 3km/h\\
 \hline
 Delay spread   &  30 \& 300ns \\
 \hline
 Modulation Coding Scheme \cite{TS0002}   &  19 \\
 \hline
 Meta task environment size $T$   &  8000 \\
 \hline
Meta training step size $\epsilon$   &  0.25 \\
 \hline
Step number per task $g$   &  32 \\
 \hline
UE number per task $\widehat{N}_{\rm ue}$   &  16 \\
 \hline
Slot number per task $\widehat{N}_{\rm slot}$   &  16 \\
\hline
 Number of groups of spatial basis $P$   &  4 \\
 \hline
 Spatial diversity degree $L_{\rm task}$   &  6 \\
 \hline
 Frequency diversity degree $M_{\rm task}$   &  6 \\
 \hline
Spatial diversity scale $\alpha$   &  0.75 \\
 \hline
Frequency diversity scale $\beta$   &  0.75 \\
 \hline
\end{tabular}}
\end{table}

\subsection{Comprehensive Performance Comparison over CDL-A/C.}
\begin{figure}[!t]
\centering
\includegraphics[scale=0.5]{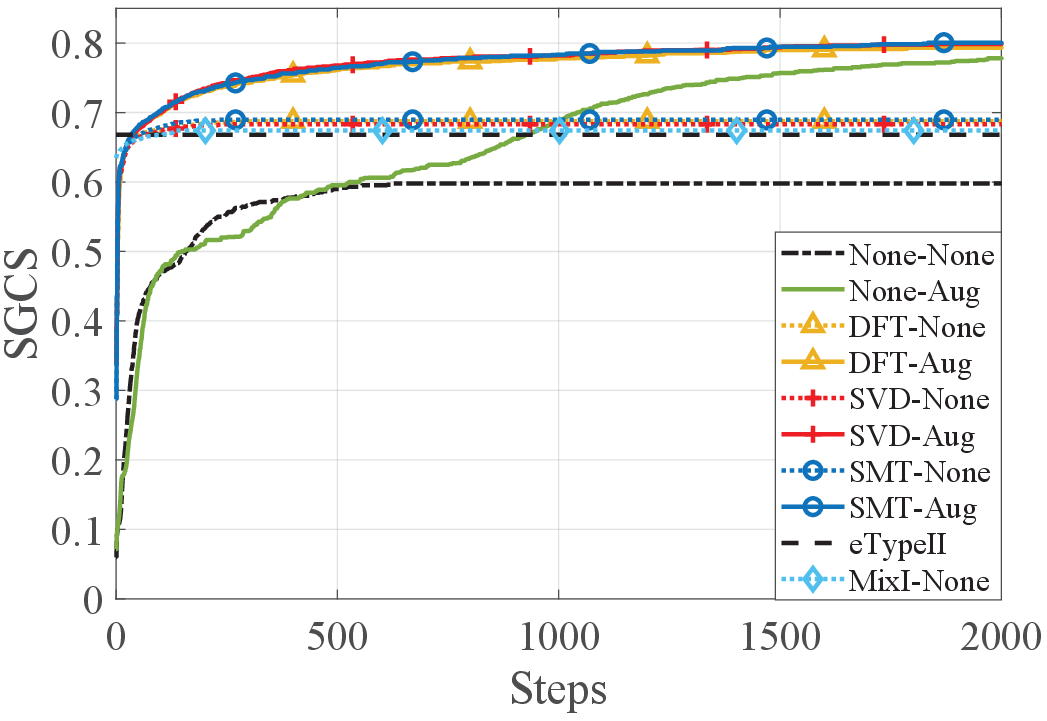}
\caption{Convergence process of target retraining phase with the number of training steps on CDL-C300 channel}
\label{fig_1}
\end{figure}

\begin{figure}[t]
\centering
\includegraphics[scale=0.5]{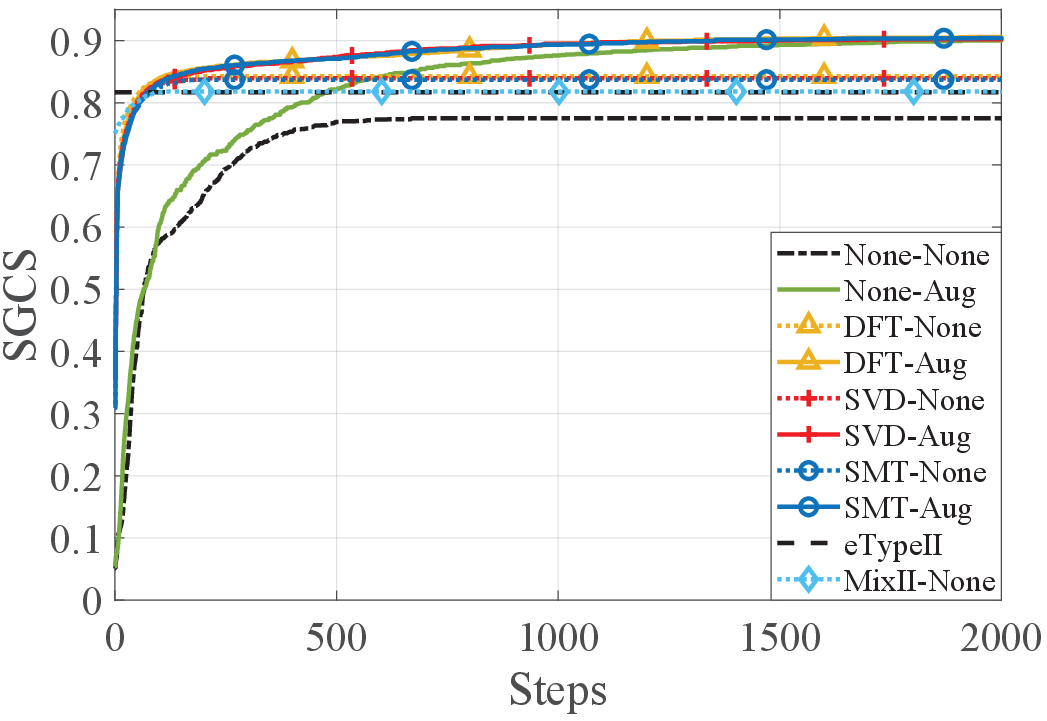}
\caption{Convergence process of target retraining phase with the number of training steps on CDL-A300 channel}
\label{fig_2}
\end{figure}

Fig. \ref{fig_1} and \ref{fig_2} show the convergence process of the target retraining phase with the number of training steps on CDL-C300 and CDL-A300 channels, respectively. Note that the vertical axis represents the best achieved SGCS on the test set within the steps. The parameter settings for knowledge-driven meta training phase are shown in Table \ref{tabSystemParameters}. Seeded CDL channels of $\widetilde{N}_{\rm ue} = 300$ UEs with $\widetilde{N}_{\rm slot} = 10$ slots and the augmented number of a UE $N_{\rm aug} = 100$ are utilized for knowledge-driven target retraining phase, therefore the data for target retraining phase is $\widetilde{N}_{\rm ue}N_{\rm aug} = 3 \times 10^{4}$. In terms of convergence speed, it can be noticed that the proposed knowledge-driven meta training methods (DFT-None, SVD-None and SMT-None) require fewer training steps to achieve convergence than the DL-based method without meta training (None-None). Even on augmented data, the proposed method (DFT-Aug, SVD-Aug and SMT-Aug) can also fit more quickly than None-Aug. From the perspective of feedback performance, the knowledge-driven meta training brings higher SGCS since each of DFT-None, SVD-None and SMT-None outperforms None-None and MixI-None. Moreover, it can be noticed that the methods of *-Aug outperform the methods of *-None which reveals that the knowledge-driven target retraining can further effectively improve the SGCS performance. Since the proposed DFT-Aug, SVD-Aug and SMT-Aug can quickly adapt to new scenarios with a small amount of data and outperform the baselines, the superiority of the proposed scheme can be well observed. Furthermore, it should be noted that the methods of DFT-None, SVD-None and SMT-None are equivalent to each other and the methods of DFT-Aug, SVD-Aug and SMT-Aug are equivalent to each other too. Thus, those three proposed methods for constructing orthogonal basis vectors are equally effective. In the following simulation results, only the results of DFT based method are shown for sake of brevity.

\begin{figure}[t]
\centering
\includegraphics[scale=0.5]{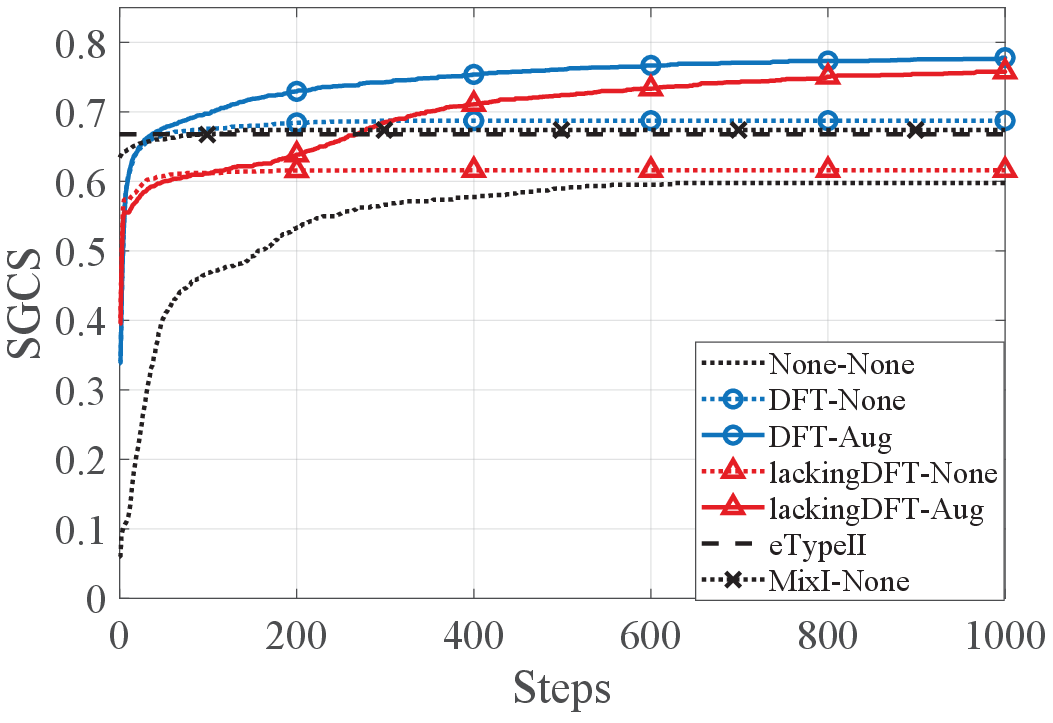}
\caption{Comparison of proposed methods with complete and incomplete basis vectors on CDL-C300 channel}
\label{fig_4}
\end{figure}

\subsection{Performance for Knowledge-Driven Meta Training Phase}
Fig. \ref{fig_4} investigates the impact of the meta task environment on the convergence and SGCS performance of target retraining phase, where the SGCS performance were evaluated with CDL-C300 channel model of $\widetilde{N}_{\rm ue} = 300$ UEs and $\widetilde{N}_{\rm slot} = 10$ slots. The augmented number of a UE $N_{\rm aug} = 100$ therefore the data for target retraining phase is $\widetilde{N}_{\rm ue}N_{\rm aug} = 3 \times 10^{4}$. Note that the method of lackingDFT-* only utilizes incomplete basis vectors with less information for meta training, i.e., (\ref{random_s_0}) and (\ref{random_f_0}) are modified by $\widehat{\mathcal{S}}_{j} = \mathrm{rand}(\{1,...,\widetilde{S}\}, L_{\rm task})$ and $\widehat{\mathcal{F}}_{j} = \mathrm{rand}(\{1,...,\widetilde{F}\}, M_{\rm task})$, respectively, where $\widetilde{S} = 6 < N_{\rm h}N_{\rm v}$ and $\widetilde{F} = 6 < N_{\rm sb}$. It can be noticed that lackingDFT-Aug using incomplete basis vectors requires more training steps to exceed eTypeII than DFT-Aug using complete basis vectors. Moreover, excluding the influence of augmentation of knowledge-driven target retraining, the incomplete basis vectors cause a 7\% loss of SGCS performance in the method of lackingDFT-None in comparison with the method of DFT-None. The proposed DFT-None also  outperforms the MixI-None in terms of the SGCS performance and the convergence speed achieving eTypeII, while the MixI-None leads to the cost of data collection. These can well reveal the rationale of Theorem \ref{theorem_2} of the proposed knowledge-driven meta training phase.

\begin{table}[b]
\centering
\caption{The number of steps required to achieve eTypeII}
\label{reach_etypeII_a}
\setlength{\tabcolsep}{0.2mm}{
\begin{tabular}{|c|c|c|c|c|c|}
\hline
\diagbox{$\widetilde{N}_{\rm ue}$}{Steps}{Schemes} & None-None& MixI-None & None-Aug & DFT-None & DFT-Aug \\ \hline
50            & /  & /     &  1093    & /     & \textbf{44}      \\ \hline
100           & /  & /      &  875    & /     & \textbf{42}      \\ \hline
150           & /  & /      &   853     & /    & \textbf{39}      \\ \hline
200           & /  & 190      &  711      & 113      & \textbf{39} \\ \hline
250           & /  & 126     & 898      & 68       & \textbf{37}      \\ \hline
300           & /  &108      & 929      & 51       & \textbf{38}      \\ \hline
\end{tabular}}
\begin{flushleft}
Note 1: the number of seeded slots per UE is fixed by $\widetilde{N}_{\rm slot} = 10$.\\
\end{flushleft}
\vspace{-0.5cm}
\end{table}

\begin{table}[b]
\centering
\caption{The number of steps required to achieve eTypeII}
\label{reach_etypeII_b}\setlength{\tabcolsep}{0.2mm}{
\begin{tabular}{|c|c|c|c|c|c|}
\hline
\diagbox{$\widetilde{N}_{\rm slot}$}{Steps}{Schemes} & None-None& MixI-None & None-Aug & DFT-None & DFT-Aug     \\ \hline
10                & /     & 108    &     1574   & 51       & \textbf{37} \\ \hline
20                & /      & 103   &   1107    & 42       & \textbf{39} \\ \hline
30                & 832    & 107   & 965     & 43       & \textbf{37} \\ \hline
40                & 1177   & 95   &  917     & 43       & \textbf{38} \\ \hline
50                & 1304   & 92   & 929      & 42       & \textbf{38} \\ \hline
60                & 1606   & 96   & 1432     & 43       & \textbf{37} \\ \hline
\end{tabular}}
\begin{flushleft}
Note: the number of seeded UEs is fixed by $\widetilde{N}_{\rm ue} = 300$.
\end{flushleft}
\vspace{-0.5cm}
\end{table}

Table \ref{reach_etypeII_a} and Table \ref{reach_etypeII_b} provide the numbers of target retraining steps required to achieve the same performance of eTypeII on CDL-C300 for various numbers of seeded UEs and slots, respectively, where `/' means that the method can not achieve the same performance of eTypeII. Note that augmented number of a UE $N_{\rm aug} = 100$ therefore the data for target retraining phase varies from $5\times 10^{3}$ to $3\times 10^{4}$. We can notice that the proposed knowledge-driven method DFT-Aug takes only 37 to 44 training steps to outperform eTypeII, while the methods of None-None and None-Aug without using knowledge-driven meta training require at least 711 steps. Thus, the overhead of training time is reduced by 93.8\% compared with no meta training phase. Since the proposed DFT-Aug also outperforms the existing meta learning method MixI-None for at least 54 steps, there is still 58.7\% training time overhead reduction. In summary, in comparison with the methods of DL-based None-None and MixI-None methods, the proposed knowledge-driven method can still maintain superior performance over traditional eTypeII with small overhead of training time. This can further reveal the effectiveness of proposed knowledge-driven meta training method for increasing training speed.

\subsection{Performance for Knowledge-Driven Target Retraining Phase}
\begin{figure}[t]
\centering
\includegraphics[scale=0.5]{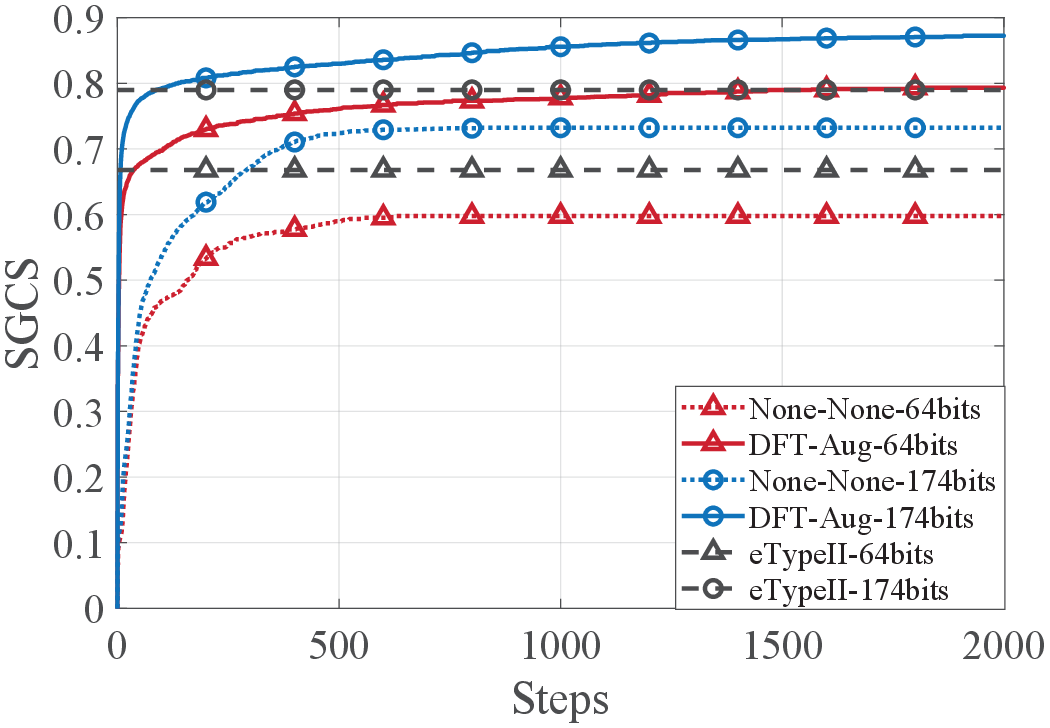}
\caption{Comparison of target retraining phase on CDL-C300 channel for different number of feedback bits}
\label{fig_3}
\end{figure}

Fig. \ref{fig_3} provide the comparison of various methods during  target retraining phase, where CDL-C300 channel is assumed and the numbers of feedback bits are $B = 64$ and $B=174$. Seeded channels of $\widetilde{N}_{\rm ue} = 300$ UEs with $\widetilde{N}_{\rm slot} = 10$ slots and the augmented number of a UE $N_{\rm aug} = 100$ are utilized for knowledge-driven target retraining phase, therefore the data for target retraining phase is $\widetilde{N}_{\rm ue}N_{\rm aug} = 3 \times 10^{4}$. It can be observed that the proposed method of DFT-Aug-* outperforms corresponding baselines for the same number of feedback bits. In addition to the excellent convergence speed and SGCS performance provided by the proposed method, we can also notice that the proposed method of 64 bits (DFT-Aug-64bits) can already achieve the same performance of eTypeII of 174 bits (eTypeII-174bits), while the overhead of feedback can be reduced by 63.2\%.

\begin{figure}[t]
\centering
\includegraphics[scale=0.5]{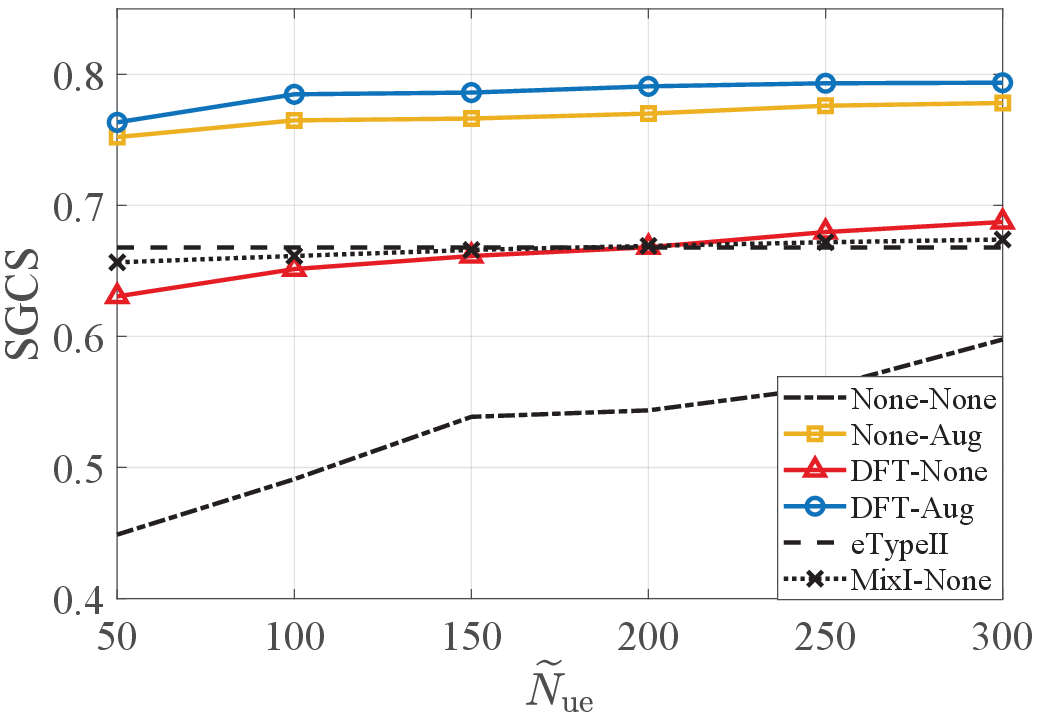}
\caption{Comparison of SGCS for varying number of seeded UEs $\widetilde{N}_{\rm ue}$ on CDL-C channel, fixing number of slots per UE $\widetilde{N}_{\rm slot} = 10$}
\label{fig_6}
\end{figure}

\begin{figure}[t]
\centering
\includegraphics[scale=0.5]{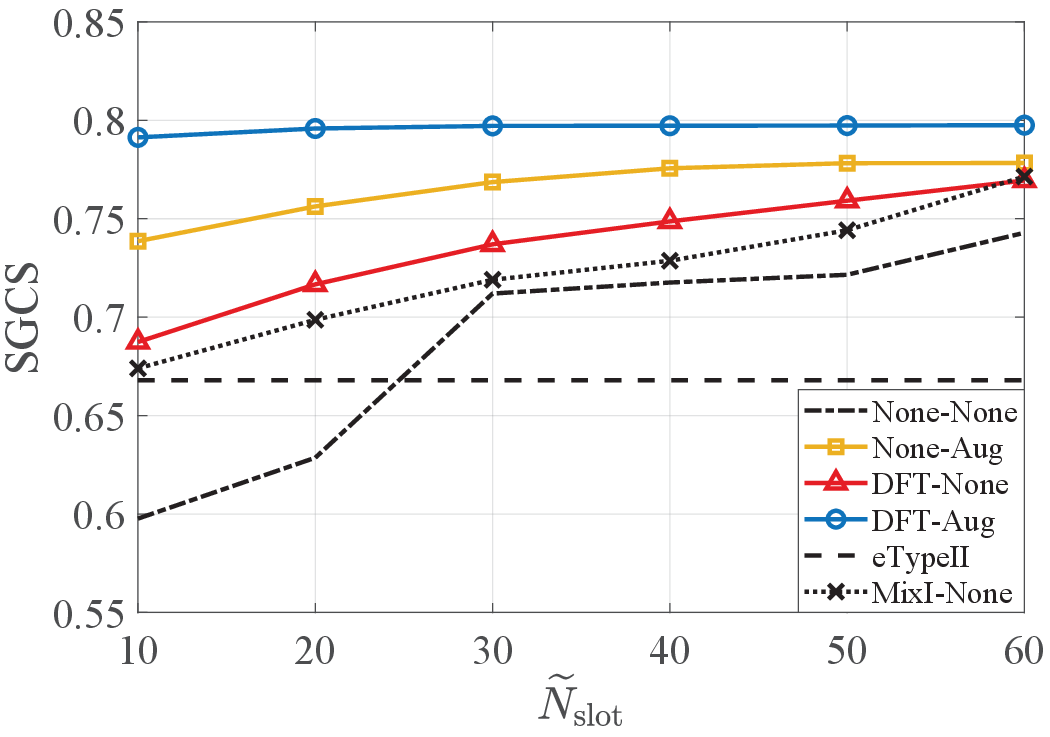}
\caption{Comparison of SGCS for varying number of slots per UE $\widetilde{N}_{\rm slot}$ on CDL-C channel, fixing number of UEs $\widetilde{N}_{\rm ue} = 300$}
\label{fig_7}
\end{figure}

To further illustrate the advantage of proposed knowledge-driven target retraining phase, Table \ref{different_aug_scheme} illustrates the SGCS performance of  the proposed method and the existing data augmentation methods \cite{xiao2022ai} for DL-based CSI feedback including noise injection, flipping, cyclic shift, random shift and rotation, where the channel of CDL-C300 of $\widetilde{N}_{\rm ue} = 300$ UEs with $\widetilde{N}_{\rm slot} = 10$ slots is utilized as seeded data. Note that except the method of flipping which can only augment to $6\times 10^{3}$ samples due to method limitation, other methods augments to $3\times 10^{4}$ samples with ${N}_{\rm aug} = 100$. It is observed that the proposed knowledge-driven method can obtain 0.1953 SGCS performance gain in comparison to the method of none augmantation. Specifically, the performance gap between proposed method and other competitors is at least 0.1526, which can demonstrate that exploiting communication knowledge can effectively bring a performance gain.

\begin{table}[b]
\centering
\caption{Comparison of different augment schemes}
\label{different_aug_scheme}
\setlength{\tabcolsep}{15mm}{
\begin{tabular}{|c|c|}
\hline
Scheme          & SGCS   \\ \hline
None            & 0.5977 \\ \hline
Noise Injection & 0.6189 \\ \hline
Flipping        & 0.6171 \\ \hline
Cyclic Shift    & 0.6404 \\ \hline
Random Shift    & 0.6225 \\ \hline
Rotation        & 0.6178 \\ \hline
Proposed        & \textbf{0.7930} \\ \hline
\end{tabular}}
\end{table}

In Fig. \ref{fig_6} and Fig. \ref{fig_7} we compare the SGCS performance training 2000 steps on different number of seeded UEs $\widetilde{N}_{\rm ue}$ and slots $\widetilde{N}_{\rm slot}$ of CDL-C channel, respectively, where ${N}_{\rm aug} = 100$ therefore the data for target retraining phase varies from $5\times 10^{3}$ to $3\times 10^{4}$. It is observed that the proposed knowledge-driven method of DFT-Aug outperforms the baselines. Specifically, the performance gaps between the methods i) DFT-None and None-None and ii) DFT-Aug and None-Aug can demonstrate the gain provided by proposed knowledge-driven meta training phase. The gaps between i) None-Aug and None-None and ii) DFT-Aug and DFT-None initimate the gain by using proposed knowledge-driven target retraining phase. Moervoer, we can notice in Fig. \ref{fig_7} that the performance of None-Aug improves as the number of slots $\widetilde{N}_{\rm slot}$ increased, while the performance of DFT-Aug stays almost unchanged, which implies that the proposed knowledge-driven target retraining phase requires fewer slots to achieve the performance ceiling when it is enhanced by proposed knowledge-driven meta training phase.

\begin{figure}[t]
\centering
\includegraphics[scale=0.5]{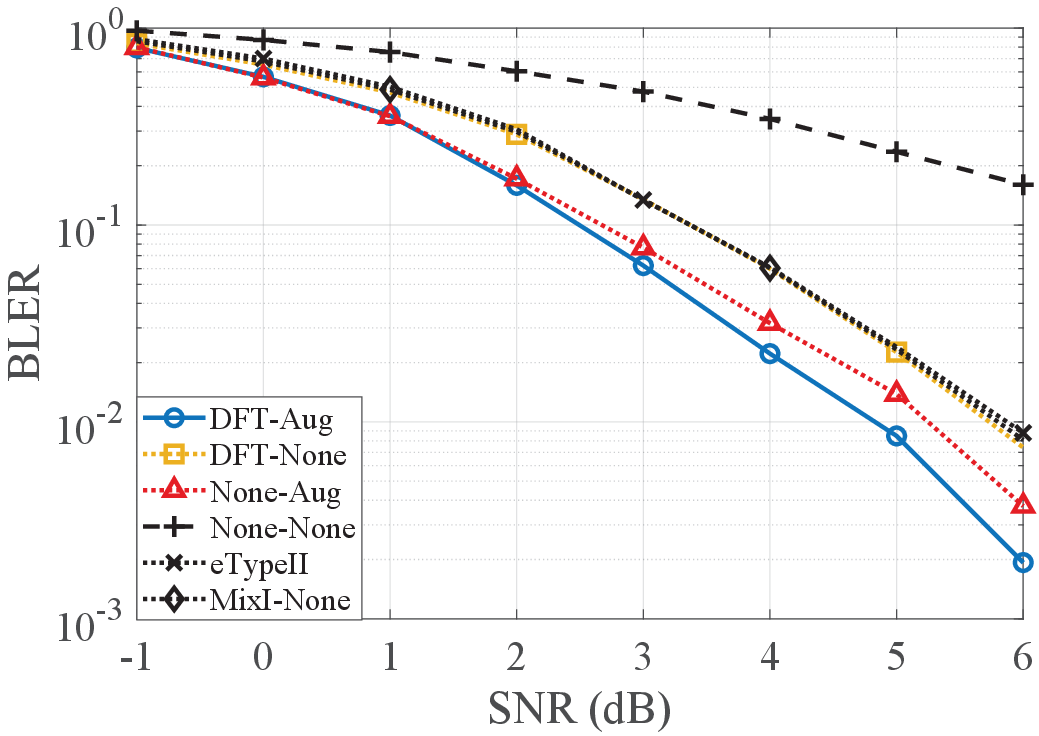}
\caption{Link-level BLER performance comparison trained on CDL-C300 for different solutions}
\label{BLER}
\end{figure}

\begin{figure}[!h]
\centering
\includegraphics[scale=0.5]{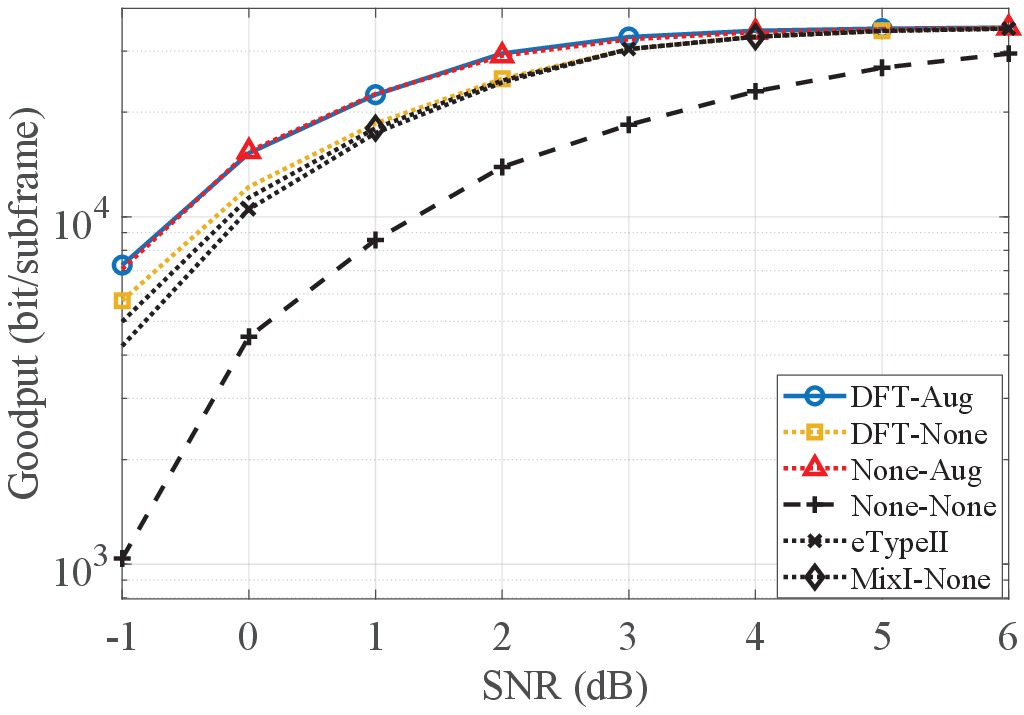}
\caption{Link-level goodput performance comparison trained on CDL-C300 for different solutions.}
\label{fig_goodput}
\end{figure}

\vspace{-0.3cm}
\subsection{Link Level Simulation}
To further prove the superiority of our proposed knowledge-driven methods, the link-level block error rate (BLER) and goodput performance is presented in Fig. \ref{BLER} and \ref{fig_goodput}, respectively, where seeded data of $\widetilde{N}_{\rm ue} = 300$ UEs with $\widetilde{N}_{\rm slot} = 10$ slots is utilized and the omnidirectional and directional antennas is deployed at UE and BS, respectively. $N_{\rm aug} = 100$ is set therefore the data for target retraining phase is $\widetilde{N}_{\rm ue}N_{\rm aug} = 3 \times 10^{4}$. LMMSE equalizer and SVD precoder is applied in simulation. The goodput measures the number of bits per subframe successfully received, and is defined by $Goodput=N_{\rm RE}\Omega\gamma(1-BLER)$, where $N_{\rm RE}$ is the number of REs forming a subframe, $\Omega$ denotes ratio of data carrying REs within a physical resource block (PRB) and $\gamma$ denotes the coderate according to MCS.
Similarly, it can be observed that the proposed method of DFT-None outperforms the method of None-None, which also proves the performance gain of knowledge-driven meta training phase. It can be notice that the DFT-None and MixI-None is comparable but the MixI-None brings higher data acquisition costs. Moreover, the performance gain of knowledge-driven target retraining phase can also be demonstrated through the performance gap between the methods of None-Aug and None-None. Since the method of DFT-Aug which jointly exploits knowledge-driven meta training and target retraining outperforms other competitors especially including eTypeII and MixI-None in terms of BLER and goodput, the performance advantages and application potential of proposed knowledge-driven approach are well demonstrated.

\begin{table}[b]
\small
\centering
\caption{FLOPs and number of trainable parameters for varing $B$.}
\setlength{\tabcolsep}{2.4mm}{
\begin{tabular}{|c|c|c|c|c|}
\hline
$B$     & FLOPs ($\times 10^{7}$)  & Trainable Parameters ($\times 10^{7}$)  \\
 \hline
$64$    & $4.277$ & 2.145 \\ \hline
$174$  & 4.295 & 2.154 \\ \hline
\end{tabular}}
\label{tabFLOPs}
\end{table}
\subsection{Complexity Evaluation}
The complexity evaluation of the proposed method is provided. For knowledge-driven meta training phase, the computational complexity for constructing CSI samples mainly lies in (\ref{csi_gen_1}) whose complexity is of the order of $O(TN_{\rm{ue}, \it{ j}}N_{\rm{slot}, \it{ j}}N_{\rm{t}}\lceil\beta M_m\rceil(\lceil\alpha L_m\rceil+N_{\rm{sb}}))$. For knowledge-driven target retraining phase, most complexity cost for constructing CSI samples comes from SVD in (\ref{R_svd}) and eigenvector deomposition in (\ref{eigdecomposition}), which have the complexity in the order of $O(F\widetilde{N}_{\rm{ue}}N_{\rm{r}}N_{\rm{delay}}N_{\rm{t}}^{3}N_{\rm{r}}^{3})$ and $O(F\widetilde{N}_{\rm ue}N_{\rm{aug}}N_{\rm{sb}}N_{\rm{t}}^{2}(N_{\rm{gran}} + N_{\rm{t}}))$, respectively. However, the spatial and computational complexity of total algorithm mainly lies in the model parameter updating in (\ref{meta_training_1}) and (\ref{target_retraining_1}), which is far beyond the above-mentioned complexity of constructing CSI samples. Therefore, the complexity evaluation of the model is further provided in Table \ref{tabFLOPs} from the perspective of floating point operations (FLOPs) and trainable parameters, which are the common metrics used to evaluate the computational and space complexity of a DL model.

\section{Standardization Potential and Prospects}\label{Standardization Potential and Prospects}
In the 3GPP work plan, starting from the 5G-advanced stage, the performance gain, scenario generalization, dataset construction, life cycle management (LCM) and potential standardization impact of DL-based solutions are to be studied. These studies will formulate the basic support for the construction of intelligent wireless communication system in 6G.

Along with those works, it can be clearly observed that there exist two bottlenecks in the development of DL-based solutions in standardization. One is the limitation of data, where it would be very challenging to construct a complete dataset for different use cases, scenarios and configurations. Another one is the limitation for scenarios. That is, how to handle the scenario adaptation DL solutions should be considered in the standardization work.

In current researches works, especially the wireless AI study of 3GPP Release 18, it is expected to achieve some engineering solutions through training data collection and transmission, on-demand switching of models and reliable LCM. However, it is meanwhile desired to try more methods to deal with the chanllenges of data and scenario bottlenecks through the breakthrough of DL technology itself, and further consider more standardization impacts, so as to pave the way for the research of AI-powered 6G.

The approach proposed in this artical explore the classical communication knowledge to solve the above challenges. When considering standardization and actual deployment issues based on the proposed method, restrictions being discussed can be relaxed on some aspects, such as the acquisition and transmission of large datasets, the latency required for online training and updates, and the storage and air interface overhead caused by model switching. Meanwhile, novel viewpoints of using communication knowledge to design DL-based method were observed. These changes will bring more diversity and space for the study of 6G intelligent system in future.

\vspace{-0.3cm}
\section{Conclusion}\label{Conclusion}
In this article, we propose a knowledge-driven meta-learning approach for DL-based CSI feedback. Specifically, instead of being trained with big dataset collected from different wireless scenarios, the meta model is trained with meta task environment constructed by intrinsic knowledge of spatial-frequency feature of CSI. Based on the meta model, one is capable of achieving rapid convergence by training on the target task dataset, which is augmented from only a few actually collected seeded data with the assistance of the knowledge of statistical feature of wireless channels. In addition, analyses are provided to explain the rationale for the improvement yielded by the knowledge. At last the simulation results are provided to demonstrate the superiority of the proposed approach from the perspective of CSI feedback performance and convergence speed.

%

\appendix
\subsection{Proof of Theorem \ref{theorem_1}}\label{app_1}
It holds that
\begin{align*}
1 &- \rho(\mathbf{W},\widehat{\mathbf{W}})\\
&= 1 - \frac{1}{N_{\rm sb}}\sum_{l=1}^{N_\textrm{sb}}\left(\frac{\|\mathbf{W}(;,\{l\})^{\rm H}\widehat{\mathbf{W}}(:,\{l\})\|_2}{\|\mathbf{W}(;,\{l\})\|_2\|\widehat{\mathbf{W}}(:,\{l\})\|_2}\right)^2\\
&\overset{(a)}{\leq} 1 - \left(\frac{1}{N_{\rm sb}}\sum_{l=1}^{N_\textrm{sb}}\frac{\|\mathbf{W}(;,\{l\})^{\rm H}\widehat{\mathbf{W}}(:,\{l\})\|_2}{\|\mathbf{W}(;,\{l\})\|_2\|\widehat{\mathbf{W}}(:,\{l\})\|_2}\right)^2 \\
&= \left(1 - \frac{1}{N_{\rm sb}}\sum_{l=1}^{N_\textrm{sb}}\frac{\|\mathbf{W}(;,\{l\})^{\rm H}\widehat{\mathbf{W}}(:,\{l\})\|_2}{\|\mathbf{W}(;,\{l\})\|_2\|\widehat{\mathbf{W}}(:,\{l\})\|_2}\right)\\
 &\cdot  \left(1 + \frac{1}{N_{\rm sb}}\sum_{l=1}^{N_\textrm{sb}}\frac{\|\mathbf{W}(;,\{l\})^{\rm H}\widehat{\mathbf{W}}(:,\{l\})\|_2}{\|\mathbf{W}(;,\{l\})\|_2\|\widehat{\mathbf{W}}(:,\{l\})\|_2}\right)\\
&\overset{(b)}{\leq} \frac{1}{N_{\rm sb}}\bar{S}\bar{F}\sigma^2 \left(1 + \frac{1}{N_{\rm sb}}\sum_{l=1}^{N_\textrm{sb}}\frac{\|\mathbf{W}(;,\{l\})^{\rm H}\widehat{\mathbf{W}}(:,\{l\})\|_2}{\|\mathbf{W}(;,\{l\})\|_2\|\widehat{\mathbf{W}}(:,\{l\})\|_2}\right)\\&\overset{(c)}{\leq}\frac{2}{N_{\rm sb}}\bar{S}\bar{F}\sigma^2
\end{align*}
where (a) holds according to the inequality of arithmetic and geometric means, (c) holds due to the fact that
\begin{align*}
\frac{1}{N_{\rm sb}}\sum_{l=1}^{N_\textrm{sb}}\frac{\|\mathbf{W}(;,\{l\})^{\rm H}\widehat{\mathbf{W}}(:,\{l\})\|_2}{\|\mathbf{W}(;,\{l\})\|_2\|\widehat{\mathbf{W}}(:,\{l\})\|_2} \leq 1
\end{align*}
and we give the proof of (b) as following.
\begin{align*}
1 &- \frac{1}{N_{\rm sb}}\sum_{l=1}^{N_\textrm{sb}}\frac{\|\mathbf{W}(;,\{l\})^{\rm H}\widehat{\mathbf{W}}(:,\{l\})\|_2}{\|\mathbf{W}(;,\{l\})\|_2\|\widehat{\mathbf{W}}(:,\{l\})\|_2}\\
&\overset{(b1)}{\leq} 1 - \frac{1}{N_{\rm sb}}\sum_{l=1}^{N_\textrm{sb}}\|\mathbf{W}(:,\{l\})^{\rm H}\widehat{\mathbf{W}}(:,\{l\})\|_2\\
&{\leq} 1 - \frac{1}{N_{\rm sb}}\left|\sum_{l=1}^{N_\textrm{sb}}\mathbf{W}(:,\{l\})^{\rm H}\widehat{\mathbf{W}}(:,\{l\})\right|\\
&=  1-\frac{1}{N_{\rm sb}}\left|\mathrm{Tr}(\mathbf{W}^{\rm H}\widehat{\mathbf{W}})\right|\\
&= 1- \frac{1}{N_{\rm sb}}\left|\mathrm{Tr}\left(\mathbf{F}\mathbf{E}^{\rm H}\widehat{\mathbf{S}}^{\rm H}\widehat{\mathbf{S}}\mathbf{E}'\mathbf{F}^{\rm H}\right)\right|\\
&\overset{(b2)}{=}1- \frac{1}{N_{\rm sb}}\left|\mathrm{Tr}\left(\mathbf{F}^{\rm H}\mathbf{F}\mathbf{E}^{\rm H}\widehat{\mathbf{S}}^{\rm H}\widehat{\mathbf{S}}\mathbf{E}'\right)\right|\\
&\overset{(b3)}{=}1- \frac{1}{N_{\rm sb}}\left|\mathrm{Tr}\left(\mathbf{E}^{\rm H}\mathbf{E}'\right)\right|\\
&= 1- \frac{1}{N_{\rm sb}}\left|\sum_{v=1}^{N_{\rm t}}\sum_{l = 1}^{N_{\rm sb}}e^{*}_{v,l}e_{v,l}\right|\\
&= 1- \frac{1}{N_{\rm sb}}\left|\sum_{v \in \mathcal{S}}\sum_{l \in \mathcal{F}}e^{*}_{v,l}e_{v,l}\right|\\
&\overset{(b4)}{=} \frac{1}{N_{\rm sb}}\left|\sum_{v \notin \mathcal{S}}\sum_{l \notin \mathcal{F}}e^{*}_{v,l}e_{v,l}\right|\\
&\leq \frac{1}{N_{\rm sb}}(N_{\rm t}-|\mathcal{S}|)(N_{\rm sb}-|\mathcal{F}|)\sigma^2\\
&=\frac{1}{N_{\rm sb}}\bar{S}\bar{F}\sigma^2.
\end{align*}
where $e_{v,l} = \mathbf{E}(\{v\},\{l\})$, (b1) holds due to the fact that $\|\mathbf{W}(;,\{l\})\|_2 = 1$ and $\|\widehat{\mathbf{W}}(;,\{l\})\|_2 \leq 1$, (b2) holds according to the property of trace, (b3) holds because of unitary matrices of $\widehat{\mathbf{S}}$ and $\mathbf{F}$, and (b4) holds because $\mathbf{E}'(\mathcal{S},\mathcal{F}) = \mathbf{E}(\mathcal{S},\mathcal{F})$ and $\mathbf{E}'(\{v\},\{l\}) = 0$ if $v \notin \mathcal{S} \parallel l \notin \mathcal{F}$. Note that $\|\mathbf{E}\|^2_{\rm F} = N_{\rm sb}$ because $\|\mathbf{W}\|^2_{\rm F} = \sum_{l=1}^{N_{\rm sb}} \mathbf{W}(:,\{l\}) = N_{\rm sb}$ and unitary matrices of $\widehat{\mathbf{S}}$ and $\mathbf{F}$. These lead to the satisfaction of (b) and completes the proof.

\vspace{-0.5cm}
\subsection{SVD based basis formulation}\label{SVD_method}

For each group of spatial orthogonal basis vector $\mathbf{S}_p, 1\leq p\leq P$, and the frequency orthogonal basis vector group $\mathbf{F}$, three full-rank random matrices $\mathbf{X}_{p}^{\rm h}\in \mathbb{C}^{N_{\rm h} \times N_{\rm h}}$, $\mathbf{X}_{p}^{\rm v} \in \mathbb{C}^{N_{\rm v} \times N_{\rm v}}$ and $\mathbf{X}^{\rm f} \in \mathbb{C}^{N_{\rm sb} \times N_{\rm sb}}$ are generated independently. Without loss of generality, we consider sampling the matrices from complex Gaussian distribution of $\mathcal{CN}(0,1)$. Then, SVD can be performed on those three matrices,
\begin{equation}\label{app_SVD_svd_1}
\begin{split}
\mathbf{U}^{\rm h}_{p},\mathbf{D}^{\rm h}_{p},\mathbf{V}^{\rm h}_{p} = \mathrm{svd}(\mathbf{X}_{p}^{\rm h})
\end{split}
\text{,}
\end{equation}
\begin{equation}\label{app_SVD_svd_2}
\begin{split}
\mathbf{U}^{\rm v}_{p},\mathbf{D}^{\rm v}_{p},\mathbf{V}^{\rm v}_{p} = \mathrm{svd}(\mathbf{X}_{p}^{\rm v})
\end{split}
\text{,}
\end{equation}
\begin{equation}\label{app_SVD_svd_3}
\begin{split}
\mathbf{U}^{\rm f},\mathbf{D}^{\rm f},\mathbf{V}^{\rm f} = \mathrm{svd}(\mathbf{X}^{\rm f})
\end{split}
\text{,}
\end{equation}
where $\mathbf{F} = \mathbf{U}^{\rm f}$ can be utilized as frequency orthogonal basis vector, $\mathbf{U}^{\rm h}_{p}$ and $\mathbf{U}^{\rm v}_{p}$ can be considered as the spatial orthogonal basis vectors corresponding to horizontal and vertical antenna ports, respectively. Therefore, the $p$-th spatial orthogonal basis vector group can be obtained by performing kronecker product, i.e.,
\begin{equation}\label{app_SVD_kron_1}
\begin{split}
\mathbf{S}_p = \mathbf{U}^{\rm h}_{p} \otimes \mathbf{U}^{\rm v}_{p}
\end{split}
\text{.}
\end{equation}

\vspace{-0.5cm}
\subsection{SMT based basis formulation}
\label{SMT based basis formulation}
The Schmidt orthogonalization can be performed on those random matrices $\mathbf{X}_{p}^{\rm h}$, $\mathbf{X}_{p}$ and $\mathbf{X}^{\rm f}$,
\begin{equation}\label{app_SMT_smt_1}
\begin{split}
\mathbf{U}^{\rm h}_{p} = \mathrm{smt}(\mathbf{X}_{p}^{\rm h})
\end{split}
\text{,}
\end{equation}
\begin{equation}\label{app_SMT_smt_2}
\begin{split}
\mathbf{U}^{\rm v}_{p} = \mathrm{smt}(\mathbf{X}_{p}^{\rm v})
\end{split}
\text{,}
\end{equation}
\begin{equation}\label{app_SMT_smt_3}
\begin{split}
\mathbf{U}^{\rm f} = \mathrm{smt}(\mathbf{X}^{\rm f})
\end{split}
\text{,}
\end{equation}
where $\mathrm{smt}(\cdot)$ denotes the procedure of Schmidt orthogonalization, $\mathbf{F} = \mathbf{U}^{\rm f}$ can be utilized as frequency orthogonal basis vector. The spatial orthogonal basis vector group can also be obtained by performing kronecker product using (\ref{app_SVD_kron_1}).

\subsection{Proof of Remark \ref{theorem_2}}\label{app_2}
Here we verify that the task using different spatial and frequency basis vectors contributes more knowledge to $\widehat{\Theta}$.
\begin{align*}
&I(\widehat{\Theta} ; \mathcal{T}_{S+1} | \mathcal{T}_{\rm sub})\\
&=H(\widehat{\Theta} | \mathcal{T}_{\rm sub}) - H(\widehat{\Theta} | \mathcal{T}_{S+1}, \mathcal{T}_{\rm sub})\\
&=\mathbb{E}\left[-\log\left(\frac{p(\widehat{\Theta} | \mathcal{T}_{\rm sub})}{p(\widehat{\Theta} | \mathcal{T}_{S+1}, \mathcal{T}_{\rm sub})}\right)\right]\\
&=\mathbb{E}\left[-\log\left(\frac{p(\widehat{\Theta}, \mathcal{T}_{\rm sub})p\left(\mathcal{T}_{S+1}, \mathcal{T}_{\rm sub}\right)}{p(\mathcal{T}_{\rm sub})p(\widehat{\Theta} , \mathcal{T}_{S+1}, \mathcal{T}_{\rm sub})}\right)\right]\\
&=\mathbb{E}\left[-\log\left(\frac{p\left(\mathcal{T}_{S+1}| \mathcal{T}_{\rm sub}\right)}{p\left(\mathcal{T}_{S+1} | \widehat{\Theta},\mathcal{T}_{\rm sub}\right)}\right)\right]\\
&=\mathbb{E}\left[-\log\left(\frac{p\left(\mathcal{S}_{S+1}, \mathcal{F}_{S+1},\mathcal{T}_{\rm sub}| \mathcal{T}_{\rm sub}\right)}{p\left(\mathcal{S}_{S+1}, \mathcal{F}_{S+1},\mathcal{T}_{\rm sub} | \widehat{\Theta},\mathcal{T}_{\rm sub}\right)}\right)\right]\\
&=-\mathbb{E}\left[-\log\left(\frac{p\left(\mathcal{S}_{S+1}, \mathcal{F}_{S+1}| \widehat{\Theta}\right)}{p\left(\mathcal{S}_{S+1}, \mathcal{F}_{S+1}\right)}\right)\right]\\
&=-\mathbb{E}\left[-\log\left(\frac{p\left(\mathcal{S}_{S+1}, \mathcal{F}_{S+1}| \widehat{\Theta}\right)p\left(\widehat{\Theta}\right)}{p\left(\mathcal{S}_{S+1}, \mathcal{F}_{S+1}\right)}\right)\right]\\
&+ \mathbb{E}\left[-\log(p(\widehat{\Theta}))\right]\\&=H(\widehat{\Theta}) - H(\widehat{\Theta} | \mathcal{S}_{S+1}, \mathcal{F}_{S+1}).
\end{align*}
Then it completes the proof.

\subsection{Proof of (\ref{h_aug}) satisfying (\ref{R_d})}\label{app_3}
It holds that
\begin{align*}
\mathbb{E}& [ \hat{\mathbf{h}}^{\rm aug}_d (\hat{\mathbf{h}}^{\rm aug}_d)^{\rm H} ]=\hat{p}_d {\mathbb{E} } [ \mathbf{U}_d \sqrt{\mathbf{D}_d} \mathbf{n} \mathbf{n}^{\rm H} \sqrt{\mathbf{D}_d^{\rm H}} \mathbf{U}_d^{\rm H}  ] \\
&=\hat{p}_d   \mathbf{U}_d \sqrt{\mathbf{D}_d} {\mathbb{E}}[\mathbf{n} \mathbf{n}^{\rm H}] \sqrt{\mathbf{D}_d^{\rm H}} \mathbf{U}_d^{\rm H}  \\
&= \hat{p}_d \mathbf{U}_d \sqrt{\mathbf{D}_d} \mathbf{I} \sqrt{\mathbf{D}_d^{\rm H}} \mathbf{U}_d^{\rm H} = \hat{p}_d \mathbf{R}_d
\text{.}
\end{align*}

\bibliographystyle{IEEEtran}
\bibliography{IEEEabrv,ref}

\newpage

%
%
%
%

\vfill

\end{document}